\documentclass[aps,prb,floatfix, twocolumn,footinbib]{revtex4}
\usepackage{color} %used for font color
\usepackage{amssymb} %maths
\usepackage{amsmath} %maths
\usepackage[utf8]{inputenc} %useful to type directly accentuated characters
\usepackage{graphicx}
\usepackage{epstopdf} % Vivek remove!

\begin{document} 
\title{A comprehensive study of electric, thermoelectric and thermal conductivities of Graphene with short range unitary and charged impurities}
\author{Vincent Ugarte, Vivek Aji, C. M. Varma} 
\affiliation{Department of Physics and Astronomy, University of California, Riverside, CA 92521} 
\begin{abstract}
Motivated by the experimental measurement of electrical and hall conductivity, thermopower and Nernst effect, we calculate the longitudinal and transverse electrical and heat transport in graphene in the presence of unitary scatterers as well as charged impurities. The temperature and carrier density dependence in this system displays a number of anomalous features that arise due to the relativistic nature of the low energy fermionic degrees of freedom. We derive the properties in detail including the effect of unitary and charged impurities self-consistently, and present tables giving the analytic expressions for all the transport properties in the limit of small and large temperature compared to the chemical potential and the scattering rates. We compare our results with the available experimental data. While the qualitative variations with temperature and density of carriers or chemical potential of all transport properties can be reproduced, we find that a given set of parameters of the impurities cannot fit all the observed data.
\end{abstract}\maketitle

\section{\label{sec:Intro} Introduction}

An unusual new electronic structure and the possibility of graphene as the basis for technologies of the future has sparked an intense effort in its fabrication and characterization. A number of spectacular properties such as conductivity in the limit of zero carrier density \cite{Novoselov11102005,PhysRevLett.99.246803,NatMater362007,Katsnelson200720}, perfect tunneling through potential barriers\cite{NatPhys822006,NatPhys212009,PhysRevLett.102.026807,Katsnelson200720} and quantum hall effect at room temperatures\cite{Novoselov02152007,Nature872009} have already been observed. While the anomalous properties of electrical conductivity has received a lot of attention, data on thermopower, Hall conductivity and Nernst\cite{PhysRevLett.102.166808,PhysRevLett.102.096807,PhysRevB.80.081413,Alexander-A.-Balandin:2008hc} have now revealed temperature and gate voltage dependence which need to be understood consistently within a single transport theory. Excellent summary of the previous theoretical and experimental work has recently become available while this work was in progress \cite{Adam20091072,2010arXiv1003.4731D}.

Graphene is a two dimensional allotrope of carbon with a hexagonal crystal structure. Since it is made up of two interpenetrating triangular sublattices, the unit cell has two atoms\cite{Wallace:1947lr}.  As long as the sublattice symmetry is preserved the two bands touch at two points in the Brillouin zone. In the vicinity of these points the hamiltonian is linear in momentum and has the structure of $\textbf{k}\cdot\vec{\sigma}$ where $\textbf{k}$ is the momentum and $\vec{\sigma}$ $=$ $\{ \sigma^{x}$, $\sigma^{y}$, $\sigma^{z } \}$\cite{Novoselov02152007,NatPhys822006}. The Pauli matrices represent pseudospin  with the two components referring to the two sublattices. The linear dispersion means that the electrons near these points in the Brillouin zone behave like relativistic massless particles in the absence of impurities and are called Dirac points\cite {Novoselov02152007,NatPhys822006}. For pure graphene the fermi surface is at the Dirac point and the density of states depends linearly on energy near the chemical potential. The vanishing density of states, the conservation of the operator $\textbf{k}\cdot\vec{\sigma}$ and the existence of two zeroes (or equivalently) valleys in the band structure are responsible for a number of novel phenomena in graphene. 

The most striking observation is that the electrical conductivity varies linearly with carrier density when the carrier density is not too small and that it is nonzero even when the carrier density goes to zero\cite{Novoselov11102005,PhysRevLett.99.246803,NatMater362007,Katsnelson200720}. Numerous attempts at explaining the latter has led to a number of different values for the minimum conductivity\cite{PhysRevB.79.125427,PhysRevB.78.085416,PhysRevLett.98.186806,JPSJ.67.2421,JPSJ.71.1318,PhysRevB.73.245403,PhysRevB.73.125411,Adam20091072,2010arXiv1003.4731D,PhysRevB.77.125409,Katsnelson:2006fr}. The reason for the different predictions can be traced to sensitivity of the results to the different approximation schemes and order of limits employed in the calculations. For example taking the zero frequency limit before the zero temperature limit does not commute with the limits taken in the opposite order. Introducing an additional scale, such as the scattering rate further complicates the order of limits providing a wide spread of possible values. Experimentally it is clear that the observed minima is not universal. Within a Boltzmann transport formalism a scattering rate inversely proportional to the energy can account for the observed linear dependence with respect to carrier density. One possible source of such scattering is long range Coulomb scatterers \cite{Adam20091072,2010arXiv1003.4731D,PhysRevLett.104.076804,PhysRevB.77.125409}. While this theory works well for finite densities, the finite minimum conductivity requires new physics near the Dirac point. Based on the observation of charge inhomogeneity in this limit\cite {NaturePhys542008,NatPhys02042008,wang:043121,ACS030332009}, a possible resolution is that the Coulomb scatterers promote the formation of charge puddles. These puddles mask the approach to zero carrier density and provide an effective mechanism for minimum conductivity. Another possibility which we work out is due to the fact that for small charge densities, the effect of Coulomb scattering due to point charged defects also needs to be calculated self-consistently. 

Alternatively, a mechanism that can provide a similar dependence of scattering rate on energy is strong scatterers in the unitarity limit \cite{PhysRevB.73.125411, schmittrink}. Within this approach, the scatterers introduce resonances and, in the independent scattering approximation, an effective impurity band forms which provides a finite density of states in the vicinity of the node. The width of this impurity band is set by the density of scatterers and for energies larger that the impurity bandwidth the linear density of states is recovered. Crucially the same parameter, i.e. density of impurities sets both the band width and the scattering rate. Qualitatively a constant conductivity at low densities crossing over to a linear in carrier density behavior is expected. A similar result occurs for Coulomb scatters as well, since for low carrier densities, even weak impurity potentials can induce resonances at low energies as the density of states goes to zero. 

In this paper, we provide the dependence of longitudinal and hall conductivity, thermopower, thermal conductivity and Nernst coefficient for various regimes in temperature, scattering rate and chemical potential for unitary scatterers as well as calculate the self-consistent scattering rate for Coulomb scatterers so as to compare the two cases. For a finite impurity bandwidth, the leading contribution to conductivity at the node is indeed universal in the limit of zero temperature. For temperatures smaller than the impurity bandwidth, the correction to the universal value is of order $(T\tau/\hbar)^{2}$, where $\tau$ is the mean free path. We compare our results to the only data available where conductivity, Hall resistance, thermopower and Nernst were measured on the same sample. Our chief conclusion is that no single choice of impurity concentration can account for all the observed data. For unitary impurities, none of the data can be fit in the entire range of gate voltages measured. For charged scatterers, the longitudinal conductivity can be fit but not the thermopower or Nernst. Since the experimental data is obtained from a two probe measurement, the fact that the Hall resistance cannot be reproduces is not surprising.

In section II, we derive the form of the impurity average self energy and discuss the nature of the effective dispersion and mean free path. A similar analysis for long range Coulomb impurities is presented in section III. The transport formalism used to derive the conductivities is discussed in section IV. The form of electrical conductivity and hall resistance is analyzed in section V. Section VI discusses the nature of thermoelectric properties. In section VII, we present the results for thermal transport. We compare our results for conductivity, hall coefficient, thermopower and Nernst with experimental data in section VIII. The details of all calculations are available in the appendix. Many parts of our work have already appeared in separate works of many others as we note in the References; our contribution is primarily the comprehensive calculations and comparison with experiments of diverse transport properties. We present asymptotic analytic expressions for the various transport quantities in a set of tables; these may be especially useful since they readily provide a physical basis for the experimental results.

\section{ Impurities Self Averaging Formalism and Carrier densities}

The Hamiltonian for graphene in tight binding formalism is\cite{Wallace:1947lr,PhysRevB.73.245403}

\begin{eqnarray}
H_{0}  &=&-t\sum_{<i,j>} (a_{i}^{\dagger}b_{j} + b_{j}^{\dagger}a_{i})\
\end{eqnarray}

\noindent Where $t$ is the nearest neighbor hopping amplitude and is related to the fermi velocity by $v_{F}=\frac{3}{2}\frac{ta}{\hbar}$. The operators in the Hamiltonian $\{a_{i}^{\dagger},b_{j}^{\dagger}\}$ represent electron creation operators on sites $i$ and $j$ in the graphene's honeycomb lattice which belong to the A and B sublattice respectively. The two atoms per unit cell leads to a $2 \times 2$ matrix for the Green's function of graphene. The two bands touch at two points in the Brillouin zone labelled by $\textbf{K}$ and $\textbf{K}'$. In the vicinity of these points the Greens function is\cite{PhysRevB.73.245403}

\begin{eqnarray}
G_{\sigma} (\overrightarrow{k},i\omega)  &=& \frac{1}{2}\sum_ {\lambda = \pm1}\frac {{ \left( {\begin{array}{cc} 1 & \lambda e^{i\Theta (\overrightarrow{k})}  \\
 \lambda e^ {-i\Theta (\overrightarrow{k})} & 1  \\
\end{array} } \right)\ }} {i\omega -\lambda |\phi (\overrightarrow{k})| }\
\end{eqnarray}

\noindent Where the function $\Theta (\overrightarrow{k})$ is equal to $-\frac{\pi}{6}+ arg(k_{x}+ik_{y})$ and the dispersion relation at the node is given by $\phi(\overrightarrow{k})=\pm \hbar v_{F}|\overrightarrow{k}|$. In the presence of impurities we have an additional term in the Hamiltonian given by

\begin{eqnarray}
H_{Imp}  &=& \sum_{<i,\sigma>}^{N_{A}} V_{i}^{A} a_{i}^{\dagger} a_{i}+\sum_{<i,\sigma>}^{N_{B}} V_{i}^{B} b_{i}^{\dagger}b_{i}\
\end{eqnarray}

In this section, the model chosen is s-wave scatter potentials in the unitary limit. If the assumption is made that we have identical impurities randomly distributed throughout the system, an impurity self-average Green's function can be used as an approximate solution to the full Green's function of graphene. The impurity self-averaging is valid if the sample size is much larger than the coherence length of electrons which is the case for most experiment. In the dilute impurity limit, the scattering of single impurities dominates transport. The self energy is calculated in the full self-consistent Born approximation (FSBA). It is important to emphasize the regime of validity of thus approximation. The FSBA is known to fail due to interference effects that involve scattering of multiple impurities \cite{aleiner, khvesch} when the chemical potential is at the Dirac point. However, at finite chemical potentials the approximation is justified as long as weak localization effects are not important\cite{peresrmp}. Further support for the FSBA approach is seen in the agreement of the density of states determined within this approximation and exact numerical methods [see \onlinecite{peresrmp} and references therein]. We use the FSBA to study transport, our results are valid for finite chemical potentials but will ultimately breakdown when the chemical potential is at or near the Dirac point. Analysis of where the crossover occurs is beyond the scope of this study.

\noindent The effect of the impurity states on transport quantities is to produce a finite value for conductivity at the node as the carrier density is no longer zero. In the dilute limit of impurities, the self energy in the full self-consistent born approximation is given by\cite {PhysRevB.73.245403}

\begin{eqnarray}
\Sigma_{FSBA} (i\omega)  &=& \frac{n_{Imp}V}{1-V\overline{G}(i\omega-\Sigma_ {FSBA}(i\omega))}\\
\overline{G}(i\omega-\Sigma(i\omega))&=&\frac{1}{N}\sum_{\overrightarrow{k}}G (i\omega-\Sigma(i\omega))\
\end{eqnarray}

\noindent For s-wave short range scatterers, the self energy is momentum independent. For Coulomb scatterers, we will assume momentum independence of the self energy. This assumption breaks down near the Dirac point\cite{PhysRevB.77.125409}. The electronic Carrier density and the change in carrier density near the node due to a change in chemical potential $\mu$ are given by

\begin{eqnarray}
\frac{n(\mu,T)}{N_{v}N_{s}}  &=& -\int \frac{d\varepsilon}{\pi}\frac {d\overrightarrow{k}} {(2\pi)^{2}} n_{F} (\varepsilon) ImG^{R}(\overrightarrow{k},\varepsilon)\\
\frac{\delta n(\mu,T)}{N_{v}N_{s}}  &=& -\int d\mu^{\prime}\int \frac{d\varepsilon}{\pi}\frac {d\overrightarrow{k}} {(2\pi)^{2}} \frac{\partial n_{F}}{\partial \mu^ {\prime}} ImG^{R}(\overrightarrow{k},\varepsilon)\nonumber\\
\end{eqnarray}

\noindent where $\{N_{v},N_{s}\}$ represent graphene's valley and spin degeneracies and $G^{R}$ are the retarded Greens function. The integral over momentum can be performed leading to

\begin{eqnarray}
n(\mu,T) &=& -\int d\varepsilon \frac{<ImG^{R}(\varepsilon)>}{\pi^{2} \hbar^{2}v_{F}^{2}} n_{F} (\varepsilon)\\
\delta n(\mu,T)  &=& -\int d\mu^{\prime}\int d\varepsilon \frac{<ImG^{R}(\varepsilon)>}{\pi^{2} \hbar^{2}v_{F}^{2}}\frac{\partial n_{F}}{\partial \mu^ {\prime}} \\
<ImG^{R}(\varepsilon)>&=& A\arctan(\frac{D^{2}+B^{2}-A^{2}} {2AB})\nonumber\\
&-& A\arctan(\frac{B^{2}-A^{2}}{2AB})\nonumber\\
&+&\frac{B}{2}\ln(\frac{(D^{2}+B^{2}-A^{2})^{2}+(2AB)^{2}} {(B^{2}-A^{2})^{2}+(2AB)^{2}}) \nonumber\\
\end{eqnarray}

\noindent where  $D$ is the electronic band width and the functions $\{A(\varepsilon),B(\varepsilon)\}$ are given by $\{ \varepsilon - Re \Sigma (\varepsilon), -Im\Sigma (\varepsilon)\}$.

\subsection{Self Energy}

\begin{figure}
\begin{center}
\includegraphics[width= \columnwidth]{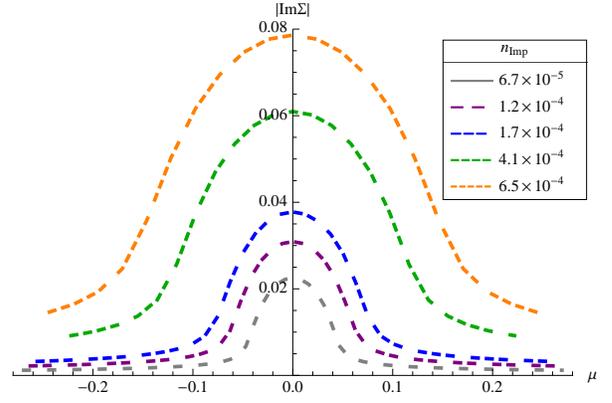}
\caption {The absolute value of the imaginary part of the self energy is plotted as a function of the energy. The impurity concentration of unitary scatterers used to find the self energy curves is given in the legend.}
\label{selfenergy}
\end{center}
\end{figure}

In fig.\ref{selfenergy} we plot $| Im \Sigma |$ as a function of energy ($\epsilon$) for several impurity concentrations. The impurity concentration is defined as the number of impurities per Carbon atom per unit area. As the impurity concentration is increased the scattering rate (which is proportional to $| Im \Sigma|$) increases. It is weakly dependent on energy near the node crossing over to a $1/\epsilon$ dependence for large energies. The crossover scale is determined by the bandwidth (D) times the square root of the impurity concentration \cite{PhysRevB.76.193401}. This crossover scale is the impurity band width. One can regard the effect of unitary scatters as producing resonances that fill in the density of states in the vicinity of the node. The linear density of states is recovered beyond the impurity band width. Unlike Born-scattering, beyond the impurity band width the scattering rate remains inversely proportional to the energy. Qualitatively, in this regime, the physics is similar to having weak Coulomb scatterers. Physical insight is gained by taking various limit of the change in electronic carrier density with respect to temperature, chemical potential and impurity band width.

\begin{eqnarray}\label{carrier-densities}
\delta n_{\mu} &\approx& \frac{sgn(\mu)|\mu|^{2}}{2\pi \hbar^{2}v_{F}^{2}}, \quad |\mu|>>|Im\Sigma(0)|>>T\\
\delta n_{T} &\approx& \frac{T\mu}{2\pi \hbar^{2}v_{F}^{2}}, \quad T>>|\mu|>>|Im\Sigma(0)|\\
\delta n_{Im\Sigma} &\approx& \frac{\mu|Im \Sigma(0) |\ln(\frac{D^{2}}{|Im\Sigma|^{2}})}{\pi^{2} \hbar^{2}v_{F}^{2}}, \quad |Im\Sigma(0)|>>|\mu|\nonumber\\
\end{eqnarray}

\noindent In the limits above, the electronic band size has been taken to be the largest energy scale and the real part of the self energy absorbed in a suitable redefinition of the chemical potential. In obtaining eqn.\ref{carrier-densities}, we have assumed that the imaginary part of the self energy is roughly constant up to the impurity band width with its value determined at zero energy. The integrals are approximated as $\int d \varepsilon \frac{\partial n_{F}(\varepsilon)}{\partial \mu}(\cdots) \to \int_{-T+\mu}^{T+\mu} \frac{d \varepsilon}{2T}(\cdots)$ and $\int d \varepsilon \frac{\partial n_{F}(\varepsilon)}{\partial \mu}(\cdots) \approx \int d \varepsilon \delta (\mu-\varepsilon)(\cdots) = (\cdots)|_{\varepsilon = \mu}$ at high and low temperatures respectively. Since $\delta n =\int d^{2}\textbf{k} \int_{0}^{\mu} d \epsilon d\textbf{k}\delta(\epsilon- \epsilon\left(\textbf{k}\right))$ we can extract an effective dispersion relation in the different regimes
\begin{eqnarray}\label{eff_disp}
 \hbar v_{F} k_{F}& \approx&|\mu|, \quad |\mu|>>|Im\Sigma|(0)>>T \\
  \frac{\hbar^{2}k_{F}^{2}}{2m^{*}}& \approx&|\mu|\nonumber\\\nonumber
  m^{*}  & \approx&\begin{array}{l l}
 \frac{|Im\Sigma|\ln(\frac{D}{|Im\Sigma|})}{2\pi \hbar^{2}v_{F}^{2}} &, \mbox{$|Im\Sigma(0)|>>|\mu|$}\\
 \frac{T}{2 v_{F}^{2}} &, \mbox {$T>>| \mu |>>|Im \Sigma(0) |$}\\  \end{array} \
 \end{eqnarray}

\noindent In the limit where $ |\mu|>>|Im\Sigma|>>T$, the Dirac dispersion relation is preserved. In the other limits, the dispersion is effectively that of a free electron with a mass determined by the dominant energy scale. Given this form we can use the effective dispersion in eqn.\ref{eff_disp} to calculate the longitudinal conductivities in various regimes\cite{ashcroft}.

\begin{eqnarray}\label{drude-cond}
\sigma  &\approx& \frac{e^{2}}{h}\frac{T} {2| Im\Sigma|}, \quad T>>|\mu|>>|Im\Sigma|\\ \nonumber
\sigma  &\approx& \frac{e^{2}}{h}\frac{|\mu|}{2| Im\Sigma|}, \quad |\mu|>>|Im\Sigma|>>T\\\nonumber
\sigma  &\approx& \frac{e^{2}}{h}\frac{2}{\pi}\ln|\frac{D}{Im\Sigma}|, |Im\Sigma|>>|\mu|\
\end{eqnarray}

\noindent For small chemical potential, we find that the conductivity is constant and depends logarithmically on the imaginary part of the self energy which in turn is proportional to the square root of the impurity concentration\cite {PhysRevB.76.193401}. Thus, within this picture, we obtain a finite minimum conductivity which is not universal. As we increase the chemical potential, at low temperatures, we crossover to a regime where the conductivity scales as $\mu^{2}\propto \delta n$ (note in this regime $Im\Sigma \propto 1/\mu$). Both these features (non-universal minimum conductivity and linear dependence on carrier density) are in qualitative agreement with observed data in graphene. There is also a constraint implicit here that the crossover scale is controlled by the same parameter that determines the value of the conductivity minimum.

\subsection{Mean Free Path}

\begin{figure}
\begin{center}
\includegraphics[width=\columnwidth]{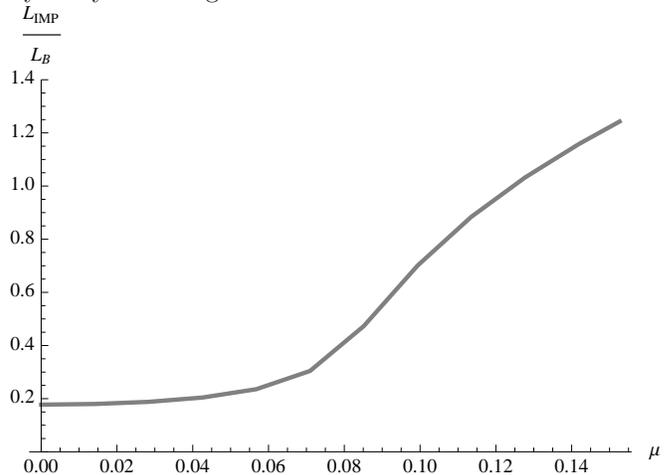}
\caption {The mean free scattering length ($L_{Imp}$) in units of cyclotron radius ($L_{B}$) is plotted as a function of the chemical potential ($\mu$). The impurity concentration of unitary scatterers is $n_{i}=2.2  \times 10^{-4}$ and the cyclotron radius is calculated in an 8 Tesla magnetic field. The Fermi velocity used is $v_{F} =1.0*10^{6} m/sec$}
\label{mfp}
\end{center}
\end{figure}

We study the transport properties in weak magnetic fields where we are in the hydrodynamic limit ($\omega_{c}\tau \ll 1$). In fig.\ref{mfp} we plot the ratio of the mean free path to the cyclotron radius as a function of the chemical potential for a fixed magnetic field. The mean free path is roughly constant up to the impurity bandwidth which in this case is 0.06 eV. Beyond this scale, the mean free path grows suggesting an energy dependent scattering rate that becomes smaller at higher energies. This behavior is consistent with the behavior of the imaginary part of the self energy. For the majority of the range shown in fig.\ref{mfp}, the mean free path is much smaller than the magnetic length for a field of 8 Tesla, for this range we are within the hydrodynamic regime.

\subsection{Specific heat} 

Thermoelectric transport coefficients, such as thermoelectric power, Nernst and thermal conductivity depend on the specific heat at constant volume. The energy density and specific heat in graphene in the presence of unitary scatterers are given by\cite{ashcroft}

\begin{eqnarray}
\frac{<E>(\mu,T)}{N_{v}N_{s}}  &=& -\int \frac{d\varepsilon}{\pi}\frac {d\overrightarrow{k}} {(2\pi)^{2}} \varepsilon n_{F} (\varepsilon) ImG_{\overrightarrow{k}}^{R}(\varepsilon)\\
\frac{c_{v}(\mu,T)}{k_{B}N_{v}N_{s}}  &=& -\int \frac{d\varepsilon}{\pi}\frac {d\overrightarrow{k}} {(2\pi)^{2}} \varepsilon \frac{\partial n_{F}} {\partial k_{B}T} ImG_{\overrightarrow{k}}^{R}(\varepsilon) \nonumber\\
\end{eqnarray}

 \begin{widetext}
\begin{center}
\begin{table}\caption{Carrier Density and Specific Heat}
\scalebox{1.2}{
    \begin{tabular} { | @{}c@{} |@{} c@{} |@{} c@{} |@{} c@{} |@{} c@{} |}
    \hline
    Quantity &$(I) T , |\mu|<< |\frac{\hbar}{2\tau}|$&$ (II)T,  |\frac{\hbar}{2\tau}|<<|\mu|$ &$(III) |\frac{\hbar}{2\tau}|<<|\mu|<<T$&$(IV) |\mu| << |\frac{\hbar}{2\tau}|<<T$ \\ \hline
    $\delta n$ & $\frac{\mu|Im \Sigma |\ln(\frac{D^{2}}{|Im\Sigma|^{2}})}{\pi^{2} \hbar^{2}v_{F}^{2}}+\cdots$ & $\frac{sgn(\mu)|\mu|^{2}}{2\pi \hbar^{2}v_{F}^{2}}+\cdots$ & $\frac{T\mu}{2\pi \hbar^{2}v_{F}^{2}} +\cdots$ & $\frac{\mu|Im \Sigma |\ln(\frac{D^{2}}{|Im\Sigma|^{2}})}{\pi^{2} \hbar^{2}v_{F}^{2}}+\cdots$  \\ \hline
    $\frac{c_{v}}{k_{B}}$ & $\frac{\pi^{2}}{3}(\frac{k_{B}T|Im\Sigma |\ln(\frac{D^{2}}{|Im\Sigma|^{2}})}{\pi^{2} \hbar^{2}v_{F}^{2}})+\cdots$ & $\frac{4\pi^{2}}{3}(\frac{k_{B}T|\mu|}{2\pi \hbar^{2}v_{F}^{2}})+\cdots$ & $\frac{1}{2}(\frac{(k_{B}T)^{2}}{2\pi \hbar^{2}v_{F}^{2}}) +\cdots$ & $\frac{1}{3}(\frac{k_{B}T|Im\Sigma |\ln(\frac{D^{2}}{|Im\Sigma|^{2}})}{\pi^{2} \hbar^{2}v_{F}^{2}})+\cdots$  \\ \hline
    \end{tabular}}\label{densitytable1}
\end{table}
\end{center}
\end{widetext}

\noindent With the same approximations used to derive the conductivity above, the specific heat and electronic carrier density, to leading order in the appropriate small parameter, in different regimes are given in table \ref{densitytable1}. The specific heat divided by temperature is proportional to $\partial\delta n/\partial\mu$. This is natural as we expect the two quantities to be proportional to the density of states which in turn is  the imaginary part of the self energy.

\section{Coulomb Scatterers}   

In this section, we consider the nature of scattering in the presence of charge impurities. We follow the same approach as in the case of unitary scatterers and compute the self energy in the self consistent Born approximation. To first order, within born approximation the self-energy due to screened Coulomb scatters has the form:

\begin{eqnarray}\label{selfenergydefinition}
\Sigma (\overrightarrow{k},i\omega_{n})&=& \frac{n_{i}}{\Omega} \sum_{\overrightarrow{k'}}|U (\overrightarrow{k'}, \overrightarrow{k},i\omega_{n})|^{2} G^{0}(\overrightarrow{k'},i\omega_{n}) \nonumber\\
\end{eqnarray}

\begin{eqnarray}
U (\overrightarrow{k'}, \overrightarrow{k},i\omega_{n})&=& \frac {(2\pi e^{2})/(\kappa \epsilon_{0})}{|\overrightarrow{k}-\overrightarrow{k'}|+q_{TF} (\overrightarrow{k'},i\omega_{n})}\nonumber\\
\end{eqnarray}

\noindent where $n_{i}$ is the concentration of charge impurities, $e$ is the charge of an electron, $\Omega$ is the area, $\epsilon_{0}$ is the vacuum permittivity, $G^{0}(\overrightarrow{k'},i\omega_{n}) $ is the green's function in the absence of impurities and $\kappa$ is the permittivity of the substrate. The function $q_ {TF} (\overrightarrow{k},i\omega_{n}) $ is the inverse Thomas Fermi screening length. In general, the self energy is a 2$\times$2 matrix and must be handled in a self-consistent manner. In order to correctly account for changes in the ground state energy the Green function in eq.(\ref{selfenergydefinition}) must be replaced by the full Green's function.

For $\overrightarrow{k}=0$ the self energy is diagonal and has the form:

\begin{eqnarray}\label{selfenergydefinitionone}
\Sigma (i\omega_{n})&=& \frac{n_{i}}{\Omega} \sum_{\overrightarrow{k'}}\left|\frac {(2\pi e^{2})/(\kappa \epsilon_{0})}{k'+q_{TF} (i\omega_{n})}\right|^{2} G_{AA}(\overrightarrow{k'},i\omega_{n}) \nonumber\\
G_{AA}(\overrightarrow{k},i\omega_{n}) &=& \frac{i\omega_{n}-\Sigma(i \omega_{n})} {(i\omega_{n}-\Sigma(i \omega_{n}))^{2}-|\phi (\overrightarrow{k}) |^{2}} \nonumber\\
\end{eqnarray}

\noindent The density of states and the inverse Thomas Fermi screening length are given by

\begin{eqnarray}\label{selfenergydefinitiontwo}
q_{TF} (i\omega_{n})&=& \frac{2\pi e^{2}}{\kappa \epsilon_{0}}\int d \varepsilon   N(\varepsilon) \frac{\partial n_{F}}{\partial \mu}\nonumber\\
N(\varepsilon) &=& -\frac{<ImG^{R}(\varepsilon)>}{\pi^{2} \hbar^{2}v_{F}^{2}}\\
<ImG^{R}(\varepsilon)>&=& A\arctan(\frac{D^{2}+B^{2}-A^{2}} {2AB})\nonumber\\
&-& A\arctan(\frac{B^{2}-A^{2}}{2AB})\nonumber\\
&+&\frac{B}{2}\ln(\frac{(D^{2}+B^{2}-A^{2})^{2}+(2AB)^{2}} {(B^{2}-A^{2})^{2}+(2AB)^{2}}) \nonumber\\
\end{eqnarray}

\noindent To obtain the transport coefficients in the presence of Coulomb scatterers we substitute the scattering rate obtained here in expressions derived for unitary scatterers. An important point to emphasize is that the self consistency yields a finite screening length at the node while crossing over to the inverse fermi wave vector at large carrier densities. Thus, the Coulomb potential is screened providing a mechanism for finite density of states and conductivity at zero bias. An alternative approach to obtain effective screening is to build on the observation of inhomogeneities near the Dirac point\cite{NatPhys02042008}. The real space density fluctuations lead to screening that can be determined by solving for the Thomas-Fermi screening length self consistently given a random potential distribution\cite{2007PNAS..10418392A,polini}. Determining whether the impurity states obtained via our self consistent approach is equivalent to the effective inhomogeneous electron gas in real space requires is beyond the scope of this article.

The self energy is momentum independent at the node. We assume that vertex corrections are small and compute the self energy considering the mass-shell only. While this is exact for unitary scatterers and for large chemical potentials for Coulomb scatterers, the nature of momentum dependence of the vertex correction near the node for the latter case is unclear. While a sizable momentum dependence has been calculated \cite{PhysRevB.77.125409}, the study ignores the effect of screening due to the formation of mid gap states. Our approach is to solve for the effective screening determined by the self energy which renders the approximation of momentum independence applicable. For a finite impurity concentration eq.(\ref{selfenergydefinitionone}) is simplified to the form:

\begin{eqnarray}\label{selfenergydefinitionthree}
Im\Sigma (0)&=& \frac{n_{i}}{\Omega} \sum_{\overrightarrow{k}}\left|\frac {(2\pi e^{2})/(\kappa \epsilon_{0})}{k+q_{TF} (0)}\right|^{2} \frac{-Im\Sigma(0)} {Im\Sigma^{2}(0)+|\phi (\overrightarrow{k}) |^{2}} \nonumber\\
\end{eqnarray}

\begin{figure}
\begin{center}
\includegraphics[width= \columnwidth]{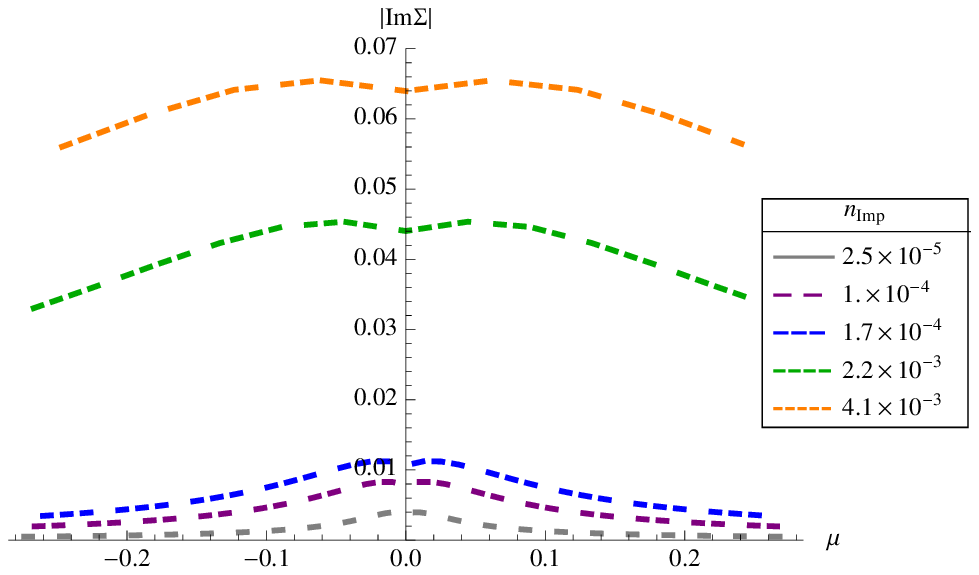}
\caption {The absolute vale of the imaginary part of the self energy is plotted as a function of the energy. The impurity concentration of Coulomb scatterers used to find the self energy curves is given in the legend.}
\label{chselfenergy}
\end{center}
\end{figure}

\noindent The summation over momenta can be performed in closed form, but the solution can only be obtained numerically. The screening obtained at zero energy allows us to approximate the self energy to be independent of momentum and compute its dependence on the chemical potential. The results are shown in fig.\ref{chselfenergy}. We have chosen impurity concentrations that yield the best fit to the data analyzed. The only free parameter is the distance of the impurities from the graphene sheet. Unlike unitary scatterers, which are part of the graphene layer itself, the Coulomb scatterers are in the substrate. In these calculations, we have assumed that the charged impurities are on the graphene sheet, thus requiring a very small concentration to fit the data.

Qualitatively the imaginary part of the self energy is similar to that of unitary scatterers. At large carrier densities, the scattering rate falls off as $1/\mu$. This is clear in fig.\ref{chselfenergy} for smaller impurity concentrations. The divergence at zero chemical potential is cutoff by the emergence of a finite screening length. The finite scattering rate is responsible for the observed minima in conductivity within this scenario. The existence of charge inhomogeneities at low bias has been experimentally observed and does lead finite conductivity at the Dirac point. Whether the real space realization of this phenomena yields charge puddles is an open question that is beyond the scope of this work. 

\section{Transport Formalism}

In response to applied electromagnetic fields and thermal gradients, the electrical and heat current are induced. Within linear response formalism, these quantities are related as\cite {ashcroft,nla.cat-vn1437043}

\begin{eqnarray}
	\overrightarrow{G} &=&  \overrightarrow{E} + \overrightarrow{\nabla}(\frac{\mu}{e})\\ \nonumber
	\overrightarrow{J} &=&  (\sigma^{(0)})(\overrightarrow{E}) + (\beta^{(1)})(- \frac{\overrightarrow{\nabla}(T)}{T})\\ \nonumber
	\overrightarrow{J_Q} &=&  (\beta^{(2)})(\overrightarrow{E}) + (\kappa^{(3)})(- \frac{\overrightarrow{\nabla}(T)}{T})\
\end{eqnarray} 

\noindent  where $\overrightarrow{J}$ is the charge current density, $\overrightarrow{J_Q}$ is the heat current density,  $\sigma^{(0)}$ is the electrical conductivity, $\beta^{(1)} = \beta^{(2)}$ is the thermoelectric conductivity and $\kappa^{(3)}$ is the heat conductivity. The heat current density is related to the charge current density : $\overrightarrow{J_Q}= \overrightarrow{J_E }-\mu\overrightarrow{J}$, where $\overrightarrow{J_E }$ is the energy current density. Each of these conductivity tensors is computed using retarded current-current correlation function within the standard Kubo formalism. The current-current correlation function and current densities are given by\cite {nla.cat-vn1437043,PhysRevB.68.155114,PhysRevB.67.115131}:

\begin{widetext}
\begin{eqnarray}\label{polarizability}
	\Pi_{i,k}^{\alpha,\beta}(q,i\Omega_{m}) &=& \frac{-i}{V}\int_{0}^{\beta} d\tau e^{i\Omega_{m}\tau} <T_{\tau}j_{\alpha,i}(\tau, q )j_{\beta ,k}^{\dagger}(\tau, 0) >\\ \nonumber
	 j_{e,\alpha}(r_{i},t)&=&  \lim_{ \begin{subarray}{1} r_{i}' \to r_{i} \\ t' \to t \end{subarray}} \frac{e}{2m}((-i\overrightarrow{\nabla}_{r_{i}'}^{\alpha} -e\overrightarrow{A}_ {r_{i}'}^{\alpha})) \Psi^{\dagger}(r_{i},t)\Psi(r_{i}',t')-\lim_{ \begin{subarray}{1} r_{i}' \to r_{i} \\ t' \to t \end{subarray}} \frac{e}{2m}((-i\overrightarrow{\nabla}_{r_{i}}^{\alpha} +e\overrightarrow{A}_ {r_{i}}^{\alpha})) \Psi^{\dagger}(r_{i},t)\Psi(r_{i}',t')\\ \nonumber
j_{E,\alpha}(r_{i},t)&=&  \lim_ { \begin{subarray}{1} r_{i}' \to r_{i} \\ t' \to t \end{subarray}} \frac{1}{2m}((-i\overrightarrow{\nabla}_{r_{i}'}^{\alpha} -e\overrightarrow{A}_ {r_{i}'}^{\alpha})\frac{\partial}{\partial t}) \Psi^{\dagger}(r_{i},t)\Psi(r_{i}',t')+\lim_ { \begin{subarray}{1} r_{i}' \to r_{i} \\ t' \to t \end{subarray}} \frac{1}{2m}((-i\overrightarrow{\nabla}_{r_{i}}^{\alpha} +e\overrightarrow{A}_ {r_{i}}^{\alpha})\frac{\partial}{\partial t'}) \Psi^{\dagger}(r_{i},t)\Psi(r_{i}',t')\nonumber \\ \nonumber
\end{eqnarray}
\end{widetext}

\noindent where $j_{\alpha,i}(\tau, q)$ is the current density operator, $\alpha$ refers to the type of current density, $i$ refers to the components, $ j_{e,\alpha}(r_{i},t)$ and $j_{E,\alpha}(r_{i},t) $ are the electrical and energy current densities\cite {PhysRevB.67.115131},  $\{\psi, \psi^{\dagger}\}$ are the fermion annihilations and creation operators and $\overrightarrow{A}_ {r}$  is the vector potential. The conductivities can be related to the appropriate current current correlation function. For closed boundary conditions, the thermoelectric transport coefficients $S$, are related to the conductivities $\beta^{(1)}$ and $\sigma^{(0)}$: 
\begin{eqnarray}
S &=& \frac{\beta^{(1)}}{T\sigma^{(0)}}\\ \nonumber
S_{xx} &=& \frac{(\sigma_{xx}^{(0)})(\beta_{xx}^{(1)}) + (\sigma_{xy}^{(0)})(\beta_{xy}^{(1)})}{T*((\sigma_{xx}^{(0)})^2 + (\sigma_{xy}^{(0)})^2)}\\ \nonumber
e_y &=& \frac{(\sigma_{xx}^{(0)})(\beta_{xy}^{(1)}) - (\sigma_{xy}^{(0)})(\beta_{xx}^{(1)})}{T*((\sigma_{xx}^{(0)})^2 + (\sigma_{xy}^{(0)})^2)}\\\nonumber
\nu &=& \frac{(\sigma_{xx}^{(0)})(\beta_{xy}^{(1)}) - (\sigma_{xy}^{(0)})(\beta_{xx}^{(1)})}{B*T*((\sigma_{xx}^{(0)})^2 + (\sigma_{xy}^{(0)})^2)}\
\end{eqnarray}

\noindent where $S_{xx}$  is the thermopower and $e_y$ is the Nernst\cite {ashcroft,nla.cat-vn1437043}. The Nernst coefficient $\nu$ is defined similarly to the hall coefficient, but here the important quantity is the ratio of transverse electric field to longitudinal temperature gradient. Similarly, the thermal transport coefficients can be obtained from\cite {ashcroft,nla.cat-vn1437043}
\begin{eqnarray}
K &=& \frac{\kappa^{(3)}}{T}-\frac{\beta^{(2)} (\sigma^{(0)})^{-1} \beta^{(1)}}{T}\nonumber\\
\end{eqnarray}

\noindent The components can be related to Nernst and thermopower:
 
\begin{eqnarray}
K_{xx}  &=& \frac{\kappa_{xx}^{(3)}}{T}+ \beta_{xy}^{(1)}e_y - \beta_{xx}^{(1)}S_{xx}\\ \nonumber
K_{xy}  &=& \frac{\kappa_{xy}^{(3)} }{T}- \beta_{xx}^{(1)}e_y - \beta_{xy}^{(1)} S_{xx}\
\end{eqnarray}

\noindent In metals the thermoelectric transport coefficients such as Nernst and thermopower are not large and generally do not contribute to thermal conductivity. The situation is more interesting in materials where the density of state vanishes, such as graphene and high temperature superconductors. The focus of this paper is on the anomalous dependence on gate voltage and temperature of thermopower and Nernst, the latter being orders of magnitude larger than typical metals.

\section{ Electrical conductivity}

In this section, we discuss the formalism for calculating the electrical conductivity tensor and a comparison of our results with other theoretical studies is given. The electrical conductivity tensor is calculated using Kubo formula\cite{nla.cat-vn1437043}: 
\begin{eqnarray}
	\sigma_{i,j}(q, \Omega) &=& \frac{\Pi_{i,j}^{e,e}(q,\Omega +i\delta)}{\Omega}\
\end{eqnarray}

\noindent where $\Pi_{i,j}^{e,e}$ is the current-current correlation function and the indices $e$ referring to the vertex corresponding to charge current. Using the definition of the charge current density operator in eqn.\ref{polarizability}, the resulting current-current correlation function is:

 \begin{widetext}
 \begin{eqnarray}
 \Pi_ {\alpha,\beta} ^{e,e}(r_{i},r_{f};\tau) &=& <T_{\tau}(\lim_ { \begin{subarray}{1} r_{i}' \to r_{i} \\ \tau' \to \tau \end{subarray}} \frac{e}{2m}[(-i\overrightarrow{\nabla}_{r_{i}'}^{\alpha} -e\overrightarrow{A}_ {r_{i}'}^{\alpha}) - (-i\overrightarrow{\nabla}_{r_{i}}^{\alpha} +e\overrightarrow{A}_ {r_{i}}^{\alpha})  ]\Psi^{\dagger}(r_{i},\tau)\Psi(r_{i}',\tau'))\nonumber\\
 & &(\lim_{ r_ {f}' \to r_ {f}} \frac{e}{2m}[(i\overrightarrow{\nabla}_{r_ {f}'}^{\beta} -e\overrightarrow{A}_ {r_ {f}'}^{\beta}) - (i\overrightarrow{\nabla}_{r_ {f}}^{\beta} +e\overrightarrow{A}_ {r_ {f}}^{\beta})  ]\Psi^{\dagger}(r_ {f}',0)\Psi(r_ {f},0)) >\
 \end{eqnarray}
\end{widetext}

 \noindent  The calculations are presented in the appendix. The conductivity in the presence of unitary scatterers is\cite {PhysRevB.73.125411,PhysRevB.76.193401,JPSJ.76.043711}

\begin{eqnarray}
\sigma_{xx}^{DC} &=& \frac{N_{v}N_{s}e^2}{4\pi h} \int d\varepsilon \frac{\partial n_{F}}{\partial \mu}\sigma_{xx}^{K}(\varepsilon)\\
\sigma_{xy}^{DC} &=& \frac{N_{v}N_{s}e^{3}| \overrightarrow {B}|v_{F}^{2}}{2c\pi}\int \frac{d\varepsilon}{\pi} \frac{\partial n_{F}(\varepsilon)}{\partial \mu}\sigma_{xy}^{K}(\varepsilon)\\
\sigma_{xx}^{K}(\varepsilon) &=& (1+\frac{A^2+B^2}{AB} \arctan{\frac {A}{B}})\\
\sigma_{xy}^{K}(\varepsilon) &=&  \frac{1}{8AB}(\frac{B^{2}-A^{2}}{B^{2}+A^{2}} - \frac{B^{2}+A^{2}}{2AB} \arctan{\frac{2AB}{B^{2}-A^{2}}})\nonumber\\
 &-& \frac{AB}{3(A^{2}+B^{2})^{2}}\
\end{eqnarray}

\noindent The kernels of the electrical conductivity tensor have been defined in terms of the functions $\{A,B\}=\{\varepsilon - Re\Sigma(\varepsilon),-Im\Sigma(\varepsilon)\}$ (for details see appendix A). Defining  $\sigma_{0}=\frac{e^{2}}{\pi h}$ and $w_{c}^{2}=\frac{2 e | \overrightarrow {B}| v_{F}^{2}}{c\hbar}$, we analyze the dependence of conductivity in the four regimes described earlier.

\subsection{ Longitudinal and transverse electrical conductivities}

\begin{figure}
\begin{center}
\includegraphics[width= \columnwidth]{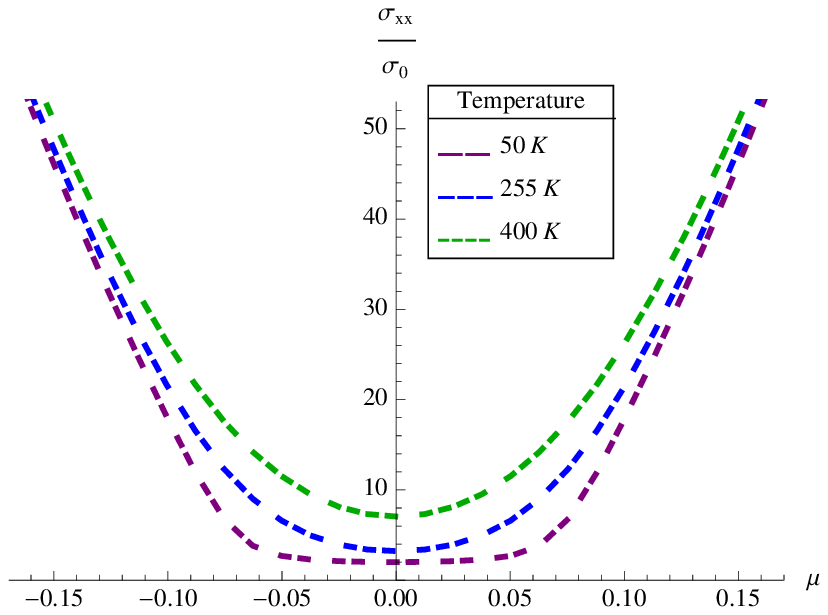}
\caption {The diagonal electrical conductivity is plotted in units of $\frac{e^2}{\pi h}$ as a function of chemical potential $(\mu)$. The impurity concentration of unitary scatterers is $n_{i}=1.7  \times 10^{-4}$, while the temperature of each curve is given in the legend.}
\label{lc}
\end{center}
\end{figure}

\begin{figure}
\begin{center}
\includegraphics[width= \columnwidth]{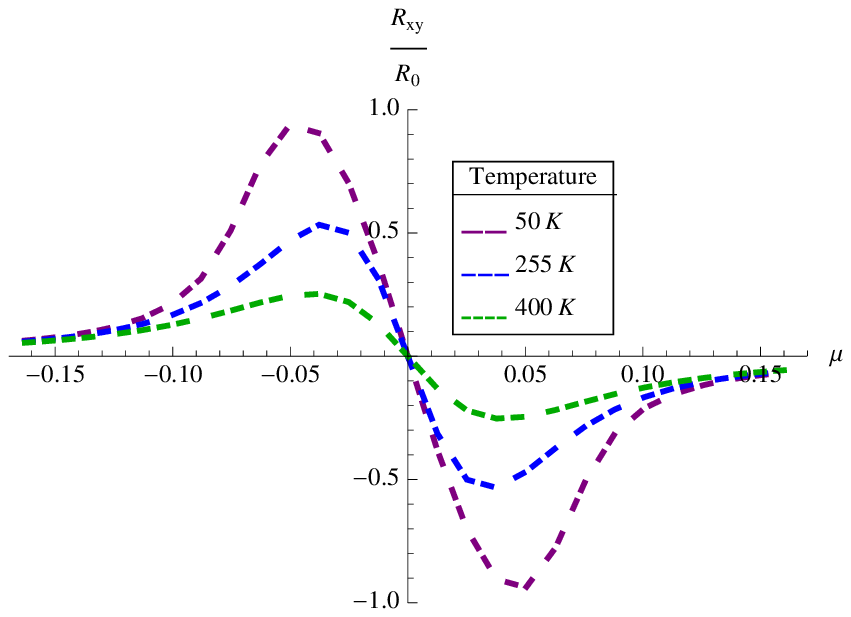}
\caption {The Hall coefficient is plotted in units of $\frac {\pi h}{|B|e^2}$ as a function of chemical potential $(\mu)$. The impurity concentration of unitary scatterers is $n_{i}=1.7  \times 10^{-4}$, while the temperature of each curve is given in the legend.}
\label{hc}
\end{center}
\end{figure}

The dependence of the longitudinal conductivity and hall resistance are shown in  fig.\ref{lc} and fig.\ref{hc}. Longitudinal conductivity has a minima at the node crossing over to a $\mu^{2}$ (linear in charge density) dependence for large carrier densities.  The crossover occurs at the impurity bandwidth. The hall coefficient is linear at the node and falls of as $1/\mu^{2}$ for large chemical potential.

\begin{widetext}
\begin{center}
\begin{table}\caption{Electrical transport coefficients}\label{etc}
\scalebox{1.4}{
\begin{tabular}{ | @{}c@{} |@{} c@{} |@{} c@{} |@{} c@{} |}
    \hline
     $Quantity $& $ (I)T, |\mu| <\left| \frac{\hbar}{ \tau}\right| $ & $(II) T,  \left|\frac{\hbar} {\tau}\right| <|\mu|$& $ (III) \left|\frac{\hbar}{\tau}\right| <|\mu| < T $ \\ \hline
    $\delta n$ & $\frac{\mu|Im \Sigma |\ln(\frac{D^{2}}{|Im\Sigma|^{2}})}{\pi^{2} \hbar^{2}v_{F}^{2}}+\cdots$ & $\frac{sgn(\mu)|\mu|^{2}}{2\pi \hbar^{2}v_{F}^{2}}+\cdots$ & $\frac{T\mu}{2\pi \hbar^{2}v_{F}^{2}} +\cdots$  \\ \hline
        $\sigma_{xx}^{DC} / \sigma_{0}$ & $2 + \frac{8}{9}(\frac{T \tau}{\hbar})^{2} + \frac{8}{3}(\frac{\mu \tau}{\hbar})^{2}+\cdots$ & $\pi|\frac{\mu\tau}{\hbar}| +\cdots$ & $\frac{\pi}{2}\frac{T\tau}{\hbar} +\frac{\pi}{4}\frac{\log{\frac {2T\tau}{\hbar}}}{\frac {T\tau}{\hbar}} +\cdots$  \\ \hline
    $\sigma_{xy}^{DC} / \sigma_{0}$ & $\frac{32(w_{c}\tau)^{2}}{3}\frac {\mu\tau} {\hbar}+ \cdots $ & $\frac{sgn(\mu)\pi(w_{c}\tau)^{2}}{2} +\cdots $ & $\frac{\pi(w_{c}\tau)^{2}}{2}\frac{\mu}{T}+\cdots$  \\ \hline
    $\tan{\Theta_{H}}$ & $\frac{16(w_{c}\tau)^{2}}{3}\frac {\mu\tau} {\hbar}+\cdots $ & $\frac{(w_{c}\tau)^{2}}{2} (\frac {\hbar} {\mu \tau}) +\cdots$ &  $(w_{c}\tau)^{2}(\frac {\hbar} {T \tau})\frac{\mu}{T} +\cdots$ \\ \hline
    $R_ {H}$ & $-\frac{8(w_{c}\tau)^{2}}{3 |B|\sigma_ {0}} \frac {\mu\tau} {\hbar}+\cdots $ & $\frac{-sgn(\mu)} {|B|\sigma_{0}}\frac{(w_{c}\tau)^{2}}{2\pi} (\frac {\hbar} {\mu \tau})^{2} +\cdots$ &  $\frac{-1} {|B|\sigma_ {0}}\frac{2(w_{c}\tau)^{2}}{\pi} (\frac {\hbar} {T \tau})^{2}\frac{\mu}{T}+\cdots$ \\ \hline
  \end{tabular}}
\end{table}
\end{center}
\end{widetext}

\noindent Analytic expressions obtained in various asymptotic limits are shown in table \ref{etc}. In the regime where $\{ T << |\frac{\hbar}{2\tau} | << \mu \}$, the longitudinal electrical conductivity is a linear function of ${\mu}/{|Im\Sigma|}$. The slope in this regime is equal to half the conductivity quanta and agrees with the other theoretical results\cite {PhysRevB.73.125411,PhysRevB.73.245411,JPSJ.76.043711}. For unitary scatterers in this regime $\tau(\mu)\sim \mu$ which implies that the conductivity is proportional to carrier density. The Hall coefficient is inversely proportionally to $({\mu}/{|Im\Sigma|})^{2}$. The coefficient of proportionality depends only on physical constants and the scattering rate. Futhermore, the Hall coefficient in terms of the carrier density is $R_{H}=-{1}/{ecn_{\mu}}$ agrees with the other theoretical results\cite {JPSJ.76.043711}.

\noindent For low carrier densities, the scattering rate is constant and the conductivity acquires a universal constant for low temperatures which is twice the quantum of conductance. The deviations are quadratic in temperature and carrier density. Rather striking is that the hall coefficient is no longer scattering rate independent and is proportional to the carrier density.

In the entropy dominated regime (III) where $\{ |\frac{\hbar}{2\tau} | << \mu << T \}$, the electrical conductivity is linear in temperature. The scale for linearity is set by the scattering rate which in this regime is inversely proportional to the chemical potential. In other words, the conductivity has a $T/\mu$ dependence at high $T$. The hall coefficient again is independent of the scattering rate. In this limit of high temperatures, for the chemical potential larger than the impurity band width, $R_{H}^{-1} = \delta n_{\mu} e c \left(T/\mu\right)^{2}$.

The hall coefficient is linear with chemical potential at low carrier densities crossing over to a $\mu^{-1}$ at large carrier densities. The crossover occurs at when the chemical potential crosses the impurity band width. As such the $R_{H}$ has a peak whose position is sample dependent but very weakly dependent on temperature. For ideal graphene, where the carrier density would be zero at the node, one would expect the hall coefficient to diverge and change from positive to negative as the chemical potential crosses zero energy. The fact that in all samples the divergence is cut off at some energy scale and the hall coefficient is zero at the node\cite{PhysRevLett.102.166808,PhysRevLett.102.096807,Novoselov11102005}. For a pure Dirac spectrum, the crossover occurs when the chemical potential crosses temperature. For unitary scatters, the crossover scale is independent of temperature for temperatures smaller than the impurity band width.

\section{ Thermoelectric transport coefficients}

The thermoelectric conductivity tensor is \cite {nla.cat-vn1437043}: 

\begin{eqnarray}
	\beta_{i,j}(q, \Omega) &=& \frac{\Pi_{i,j}^{E,e}(q,\Omega +i\delta)}{\Omega}\
\end{eqnarray}

\noindent where the $\Pi_{i,j}^{E,e}$ is the correlation function of charge and energy current densities. In the presence of an external magnetic field, the canonical momentum operators are used to define the appropriate current vertices. Since two bands touch in graphene, both electrons and holes contribute. The two contributions add for off diagonal transport but have opposite signs for thermopower. In particular, thermopower vanishes at the node and falls off as $\mu/T$ for large temperatures, in contrast to typical metals, which it would be a constant equal to the entropy per particle of a classical electron gas.

The calculation of the thermoelectric tensor is technically more complicated as the conventional Kubo formulas need to be generalized to include the effect of magnetization\cite {PhysRevB.31.7291,PhysRevB.55.2344,Smrcka-Streda:1977SS}. Fortunately the correction due to magnetization is ${cM}/{T}$\cite {PhysRevB.31.7291,JPSJ.76.043711}, which in the limit of weak magnetic field is proportional to $B^{2}$ and is neglected to leading order in magnetic field.

The kernels appearing in the calculations for thermoelectric transport are related to those that determine electrical conductivity and can be expressed as: $\beta_{\alpha\beta}^{K} = (({\mu - \varepsilon})/{e})*\sigma^{K}_ {\alpha\beta}$\cite {nla.cat-vn1437043}. For magnetic field perpendicular to the graphene sheet the kernels are (see appendix B)\cite {PhysRevB.73.125411,PhysRevB.76.193401,JPSJ.76.043711,nla.cat-vn1437043}

\begin{eqnarray}
\beta_{xx}^{DC} &=& \frac{N_{v}N_{s}e^{2}}{4\pi h} \int d\varepsilon \frac{\partial n_{F}}{\partial \mu} \beta_{xx}^{K}(\varepsilon)\\
\beta_{xy}^{DC} &=& \frac{-N_{v}N_{s}e^{3}| \overrightarrow {B}|v_{F}^{2}}{2c\pi}\int \frac{d\varepsilon}{\pi} \frac{\partial n_{F}(\varepsilon)}{\partial \mu} \beta_{xy}^{K}(\varepsilon)\\
\beta_{xx}^{K}(\varepsilon)&=&(\frac{\mu - \varepsilon}{e})(1+\frac{A^2+B^2}{AB} \arctan{\frac {A}{B}})\\
\beta_{xy}^{K}(\varepsilon)&=&(\frac{\mu - \varepsilon}{e})\{\frac{1}{8AB}(\frac{B^{2}-A^{2}}{B^{2}+A^{2}} \nonumber\\
&-& \frac{B^{2}+A^{2}}{2AB} \arctan{\frac{2AB}{B^{2}-A^{2}}}) \nonumber\\
&-& \frac{AB}{3(A^{2}+B^{2})^{2}} \} \
\end{eqnarray}

\noindent We proceed to analyze the properties in the various regimes as before. All expressions are quoted in terms $\beta_{0}=\frac {k_{B}T\sigma_{0}}{e}$, $w_{c}^{2}={2 e | \overrightarrow {B}| v_{F}^{2}}/{c\hbar}$ and $S_{0}={k_{B}}/{e}$. 

\subsection{\label{sec:model} Thermoelectric transport and Scaling Behavior}

\begin{figure}
\begin{center}
\includegraphics[width= \columnwidth] {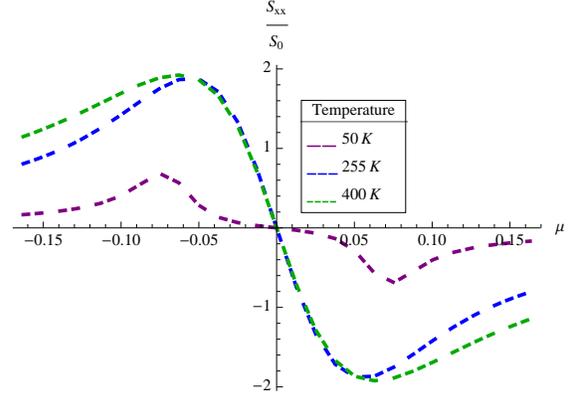}
\caption {The thermopower $(S_{xx})$ is plotted in units of $S_{0}$ as a function of chemical potential $(\mu)$. The impurity concentration of unitary scatterers is $n_{i}=1.7  \times 10^{-4}$, while the temperature for each curve is given in the legend.}
 \label{tp}
\end{center}
\end{figure}

\begin{figure}
\begin{center}
\includegraphics[width= \columnwidth] {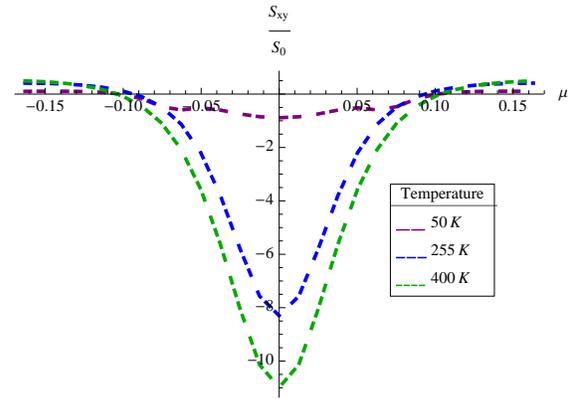}
\caption {The nernst signal $(S_{xy})$ is plotted in units of $S_{0}$ as a function of chemical potential $(\mu)$. The impurity concentration of unitary scatterers is $n_{i}=1.7  \times 10^{-4}$, while the temperature for each curve is given in the legend.}
\label{nc}
\end{center}
\end{figure}

A plot of the numerical evaluation of the thermoelectric power and Nernst signal are displayed in fig.\ref{tp} and \ref{nc}. The asymptotic dependences in the regimes identified in the previous section are shown in table \ref{ttc}.

\begin{widetext}
\begin{center}
\begin{table}\caption{Thermoelectric coefficients}\label{ttc}
\scalebox{1.3}{
    \begin{tabular} { | @{}c@{} |@{} c@{} |@{} c@{} |@{} c@{} |}
    \hline
    $ Quantity $& $ (I) T, |\mu| < \left|{\hbar\over \tau}\right| $ & $(II) T,  \left|{\hbar\over \tau}\right| <|\mu|$& $ (III) \left|{\hbar\over \tau}\right| <|\mu| < T $ \\ \hline
    $\delta n$ & $\frac{\mu|Im \Sigma |\ln(\frac{D^{2}}{|Im\Sigma|^{2}})}{\pi^{2} \hbar^{2}v_{F}^{2}}+\cdots$ & $\frac{sgn(\mu)|\mu|^{2}}{2\pi \hbar^{2}v_{F}^{2}}+\cdots$ & $\frac{T\mu}{2\pi \hbar^{2}v_{F}^{2}} +\cdots$  \\ \hline
        $\beta_{xx}^{DC} / \beta_{0}$ & $\frac{-16\pi^{2}}{9}(\frac{\tau T}{\hbar})^{2} \frac{\mu}{T} +\cdots$ & $-sgn(\mu)\frac{\pi^{3}}{3}\frac{\tau T}{\hbar}  +\cdots$ & $-\frac{\pi}{2}(\frac{T\tau} {\hbar}) +\cdots$  \\ \hline
    $\beta_{xy}^{DC} / \beta_{0}$ & $\frac {-32\pi^{2}(w_{c}\tau)^{2}}{9}\frac{\tau T}{\hbar} + \cdots $ & $sgn(\mu)\frac {\pi^{2}}{12} (w_{c}\tau)^{2} (\frac {\hbar}{\mu\tau})^{2} \frac {T}{\mu} +\cdots $ & $-\frac{\pi(w_{c}\tau)^{2}}{4}+\cdots$  \\ \hline
    $\tan{\Theta_ {H, TE}}$ & $2(w_{c}\tau)^{2} \frac {\hbar}{\mu\tau}+\cdots $ & $-\frac{(w_{c}\tau)^{2}}{4}(\frac {\hbar}{\mu\tau})^{3} +\cdots$ &  $-\frac{(w_{c}\tau)^{2}}{2}\frac {\hbar}{\tau \mu} +\cdots$ \\ \hline
    $S_ {xx}/S_{0}$ & $-\frac{8\pi^{2}}{9}(\frac {\tau T} {\hbar})^{2}\frac{\mu} {T}+\cdots $ & $-\frac{\pi^{2}}{3}\frac {T}{\mu} +\cdots$ &  $-\frac{\mu}{T}+\cdots$ \\ \hline
     $e_ {y}/S_{0}$ & $- \frac{16\pi^{2}(w_{c}\tau)^{2}}{9} \frac {\tau T}{\hbar}+\cdots $ & $\frac {\pi^{2}(w_{c}\tau)^{2}}{6}\frac {\hbar}{\mu\tau}\frac {T}{\mu} +\cdots$ &  $-\frac{(w_{c}\tau)^{2}}{2}\frac {\hbar}{\tau T}+\cdots$ \\ \hline
    \end{tabular}}
\end{table}
\end{center}
\end{widetext}

At high temperatures, the thermopower is related to the entropy per unit charge. Since both electron and hole states are thermally populated, the net charge is an imbalance between the two. At the Dirac point, the system is particle hole symmetric and the thermopower goes to zero\cite{PhysRevLett.102.166808,PhysRevLett.102.096807,PhysRevB.80.081413,Alexander-A.-Balandin:2008hc}. At small carrier densities, the difference between positive and negative charge occupations is linear in the chemical potential, and the thermopower is $\propto \mu/T$. This dependence on the chemical potential is very similar to the high temperature classical limit. The thermopower is linear in $\mu$ and the coefficient of $\mu/T$ is a measure of the relaxation time\cite{PhysRevB.76.193401,PhysRevB.80.235415,PhysRevB.73.245411}.

Consider the dependence of the thermopower on the chemical potential for temperatures smaller than the impurity band width. For small chemical potentials, the thermopower grows linearly and reaches a maximum approximately at a chemical potential of order of the impurity band width. For larger values it decreases as $T/\mu$ in agreement with the other theoretical results\cite {PhysRevB.76.193401,PhysRevB.80.235415,PhysRevB.73.245411}. As the temperature is increased and becomes larger than the impurity bandwidth, the thermopower qualitatively shows the same dependence but the peak now is at a chemical potential of order the temperature. In other words, as one increases the temperature the position of the peak in thermopower as a function of chemical potential will remain roughly constant until the temperature becomes larger than the impurity bandwidth. For larger temperatures, the peak will move to larger values of chemical potential.

For chemical potentials larger than the impurity bandwidth, the Nernst signal is proportional to the thermopower. In this regime, the scattering rate is inversely proportional to the chemical potential. Thus for unitary scatterers the ratio of the thermopower to the Nernst signal is a constant proportional to the applied magnetic field. As the chemical potential is lowered and crosses the impurity band width the two start to deviate. Within this scenario, for a fixed magnetic field, the ratio goes to zero as $\mu/\tau$. The slope of the ratio as a function of chemical potential is a direct measure of the scattering rate. Within Born approximation, the scattering rate is proportional to the density of state which for graphene is linear in energy. Thus one would expect a divergent Nernst coefficient at the node. This divergence is cutoff and the value of Nernst is proportional to $\tau^{3}$.

In the clean limit, $\tau \rightarrow \infty$, the Nernst diverges as the carrier density goes to zero\cite {PhysRevB.73.245411}. The presence of unitary scatters makes it finite and is proportional to $\tau^{3}$ at the node. The Nernst signal is still orders of magnitude larger than typical metals but the observed values of order $30 \mu V/K T$ implies a scattering rate of order $0.04 eV$ for unitary impurities. The larger the mean free path, larger is the Nernst signal and, therefore, one expects a large Nernst signal for large electrical conductivity at the node. As we will see later this is a key puzzle in graphene.

\section{ Thermal transport}

Having discussed the electrical and thermoelectric transport in previous sections, we now consider thermal transport. With the boundary condition where the electrical conductivity is zero, the three are related.

\begin{equation}
K=(\frac{1}{T})(\kappa^{(3)} -\beta ^{(2)}\frac{\beta^{(1)}}{\sigma^{(0)}})
\end{equation}
where $\kappa$ is the energy conductivity tensor which is related energy current-current correlation function.
\begin{eqnarray}
	\kappa_{i,j}(q, \Omega) &=& \frac{\Pi_{i,j}^{E,E}(q,\Omega +i\delta)}{\Omega}\
\end{eqnarray}

\noindent As in the case of thermoelectric conductivities, the effect of magnetization is of order $B^{2}$ and is dropped in our analysis. For a magnetic field perpendicular to the graphene sheet, the longitudinal and transverse components of thermal conductivity tensor are (see appendix C for details)\cite {PhysRevB.73.125411,PhysRevB.76.193401,JPSJ.76.043711,nla.cat-vn1437043}

\begin{eqnarray}
\kappa_{xx}^{DC} &=& \frac{Ne^{2}}{4\pi h} \int d\varepsilon \frac{\partial n_{F}}{\partial \mu} \kappa_{xx}^{K}(\varepsilon)\\
\kappa_{xy}^{DC} &=& \frac{-Ne^{3}| \overrightarrow {B}|v_{F}^{2}}{2c\pi}\int \frac{d\varepsilon}{\pi} \frac{\partial n_{F}(\varepsilon)}{\partial \mu} \kappa_{xy}^{K}(\varepsilon)\\
\kappa_{xx}^{K}(\varepsilon)&=&(\frac{\mu - \varepsilon}{e})^{2}(1+\frac{A^2+B^2}{AB} \arctan{\frac {A}{B}})\\
\kappa_{xy}^{K}(\varepsilon)&=&(\frac{\mu - \varepsilon}{e})^{2}\{\frac{1}{8AB}(\frac{B^{2}-A^{2}}{B^{2}+A^{2}} \nonumber\\
&-& \frac{B^{2}+A^{2}}{2AB} \arctan{\frac{2AB}{B^{2}-A^{2}}}) \nonumber\\
&-& \frac{AB}{3(A^{2}+B^{2})^{2}} \} \
\end{eqnarray}

\noindent The thermal conductivity can be expressed in terms of the thermopower and Nernst as

\begin{eqnarray}
K_{xx}^{DC}  &=& \frac{\kappa_{xx}^{DC} }{T}+ \beta_{xy}^{DC} e_y - \beta_{xx}^{DC} S_{xx}\\
K_{xy}^{DC}  &=& \frac{\kappa_{xy}^{DC} }{T}- \beta_{xx}^{DC} e_y - \beta_{xy}^{DC} S_{xx}\
\end{eqnarray}

\noindent where $K_{xx}$ is the longitudinal thermal Conductivity and $K_{xy}$ is the transverse thermal Conductivity. Given the large Nernst signal and anomalous temperature and carrier density dependence of thermopower, thermoelectric contribution to thermal conductivity is significant in graphene. Analogous to the hall coefficient we define the thermal hall coefficient,

\begin{eqnarray}
R_{T.H}  &=&\frac{-K_{xy}}{B(K_{xx}^{2}+K_{xy}^{2})}\
\end{eqnarray}

\noindent The average energy and the specific heat dependence on chemical potential, temperature and scattering rate are given in section II. We now analyze thermal transport and all results are quoted in terms of both $\kappa_{0}=\frac{\pi^{2}}{3}S_{0}\frac{k_{B}T\sigma_{0}}{e}=\frac{\pi^{2}}{3}S_{0}b_{0}$ and $S_{0}=\frac{k_{B}}{e}$.

\subsection{\label{sec:thermalcoefficient} Thermal Transport Quantities and Scaling Behavior}

\noindent The asymptotic behavior of diagonal and off diagonal heat transport are given in table \ref{httable}. 

\begin{widetext}
\begin{center}
\begin{table}\caption{Thermal transport}\label{httable}
\scalebox{1.1}{
    \begin{tabular} { | @{}c@{} |@{} c@{} |@{} c@{} |@{} c@{} |@{} c@{} |}
    \hline
    Quantity &$(I) T , |\mu|<< |\frac{\hbar}{2\tau}|$&$(II) T,  |\frac{\hbar}{2\tau}|<<|\mu|$ &$(III) |\frac{\hbar}{2\tau}|<<|\mu|<<T$&$ (IV)  |\mu| << |\frac{\hbar}{2\tau}|<<T$ \\ \hline
    $\delta n$ & $\frac{\mu|Im \Sigma |\ln(\frac{D^{2}}{|Im\Sigma|^{2}})}{\pi^{2} \hbar^{2}v_{F}^{2}}+\cdots$ & $\frac{sgn(\mu)|\mu|^{2}}{2\pi \hbar^{2}v_{F}^{2}}+\cdots$ & $\frac{T\mu}{2\pi \hbar^{2}v_{F}^{2}} +\cdots$ & $\frac{\mu|Im \Sigma |\ln(\frac{D^{2}}{|Im\Sigma|^{2}})}{\pi^{2} \hbar^{2}v_{F}^{2}}+\cdots$  \\ \hline
    $\frac{c_{v}}{k_{B}}$ & $\frac{\pi^{2}}{3}(\frac{k_{B}T|Im\Sigma |\ln(\frac{D^{2}}{|Im\Sigma|^{2}})}{\pi^{2} \hbar^{2}v_{F}^{2}})+\cdots$ & $\frac{4\pi^{2}}{3}(\frac{k_{B}T|\mu|}{2\pi \hbar^{2}v_{F}^{2}})+\cdots$ & $\frac{1}{2}(\frac{(k_{B}T)^{2}}{2\pi \hbar^{2}v_{F}^{2}}) +\cdots$ & $\frac{1}{3}(\frac{k_{B}T|Im\Sigma |\ln(\frac{D^{2}}{|Im\Sigma|^{2}})}{\pi^{2} \hbar^{2}v_{F}^{2}})+\cdots$  \\ \hline
    $\kappa_{xx}^{DC} / \kappa_{0}$ & $2 +\frac{8}{9}(\frac{T \tau}{\hbar})^{2}+ \frac{8}{3} (\frac{\tau \mu}{\hbar})^{2} +\cdots$ &$\pi \frac{\tau |\mu|}{\hbar} +\cdots$ & $\frac{3}{4 \pi}(\frac{\tau T}{\hbar}) +\cdots$ & $\frac{2}{\pi^{2}} + \frac{8}{\pi^{2}} (\frac{\tau \mu}{\hbar})^{2}  +\cdots$  \\ \hline
    $\kappa_{xy}^{DC} / \kappa_{0}$ &  $ \frac {32(w_{c}\tau)^{2}}{3}\frac{\tau \mu}{\hbar} +\cdots$ &$\frac{\pi}{2}(w_{c}\tau)^{2}sgn(\mu) +\cdots$ & $-\frac{\mu}{T}\frac{3}{2\pi}(w_{c}\tau)^{2} +\cdots$ & $\frac{32(w_{c}\tau)^{2}}{3\pi^{2}}\frac{\tau \mu}{\hbar}  +\cdots$  \\ \hline
    $\tan{\Theta_ {H, T}}$ & $\frac{16(w_{c}\tau)^{2}}{3}\frac{\tau \mu}{\hbar} +\cdots$ &  $\frac{(w_{c}\tau)^{2}}{2}\frac {\hbar}{\tau \mu}+\cdots $ & $-2(w_{c}\tau)^{2}\frac {\hbar}{\tau T} \frac {\mu}{T} +\cdots$ & $\frac{16(w_{c}\tau)^{2}}{3}\frac{\tau \mu}{\hbar}  + \cdots$ \\ \hline
    $R_ {H,T}$  & $-\frac{8}{3}\frac{(w_{c}\tau)^{2}} {|B|k_{0}}\frac{\tau \mu}{\hbar} +\cdots$ &$ -\frac{1}{2\pi}\frac{(w_{c}\tau)^{2}} {|B|k_{0}}(\frac {\hbar}{\tau \mu})^{2}sgn(\mu)+\cdots $ &  $-\frac{8 \pi(w_{c}\tau)^{2}} {3 |B|k_{0}}(\frac {\hbar}{\tau T})^{2} \frac {\mu}{T} +\cdots$ & $-\frac{8 \pi^{2}(w_{c}\tau)^{2}} {3|B|k_{0}}\frac{\tau \mu}{\hbar} + \cdots$ \\ \hline
    \end{tabular}}
\end{table}
\end{center}
\end{widetext}

\begin{figure}
\begin{center}
\includegraphics[width= \columnwidth] {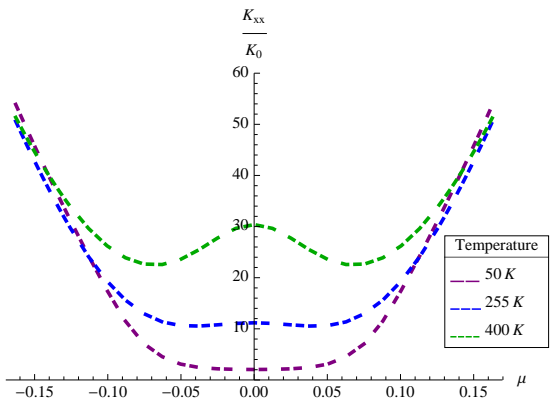}
\caption {The thermal conductivity $(K_{xx})$ is plotted in units of $\kappa_{0}$ as a function of chemical potential $(\mu)$. The impurity concentration of unitary scatterers is $n_{i}=1.7  \times 10^{-4}$, while the temperature for each curve is given in the legend.}
\label{ht}
\end{center}
\end{figure}

\begin{figure}
\begin{center}
\includegraphics[width= \columnwidth]{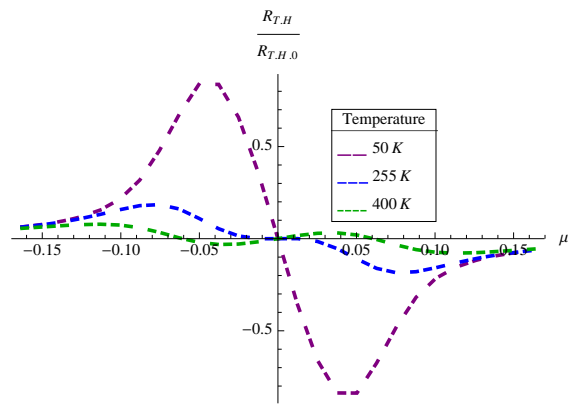}
\caption {The thermal hall coefficient $(R_{T.H})$ is plotted in units of $R_{0,T,H} = 1/B \kappa_{0}$ as a function of chemical potential $(\mu)$. The impurity concentration of unitary scatterers is $n_{i}=1.7  \times 10^{-4}$, while the temperature for each curve is given in the legend. In regimes $I$, $II$ and $III$ it is proportional to the hall coefficient with the constant of proportionality being $\sigma_{0}/\kappa_{0} =\pi^{2}k_{B}^{2}T/3e^{2}$.}
\label{odht}
\end{center}
\end{figure}

Numerical results for thermal transport are displayed in fig.\ref{ht} and fig. \ref{odht}. The thermal conductivity depends on the correlations of the energy current as well as thermopower and Nernst. The anomalous behavior of the thermoelectric coefficients in graphene has a dramatic effect on heat transport. In particular Wiedemann-Franz law is not universally obeyed. At low temperature thermal conductivity is qualitatively similar to electrical conductivity and corrections are of order $(T\tau)^{2}$. However, at the node and for very large carrier densities, the Wiedemann-Franz law is obeyed. In the intermediate regime, the deviation grows as the temperature approaches the impurity bandwidth. In particular a peak develops at the node as the temperature is increased and becomes of order the impurity bandwidth.

\section{Comparison of Experiments Data and Numerical Results}\label{densitytable}

\subsection{Unitary Scatterers}

We compare the transport dominated by unitary scatterers with experimental data on graphene\cite {PhysRevLett.102.166808}. Since much of the data is obtained as a function of gate voltage and temperature, we need to determine the dependence of chemical potential on gate voltage. Experimental control over carrier concentration is achieved by using a parallel plate geometry. For the experimental range of parameters used, we can assume that the capacitance of the device is constant which implies that the gate voltage is linearly proportional to the number of charge carriers: $Q = C V_{g}$.

The only fitting parameter is the impurity concentration. Other relevant parameters are: (1) band width ($D= 9.6 eV$), (2) Lattice constant ($a=1.42 \mathring{A}$) and fermi velocity ($v_{F}= 1.0 \times 10^{6}$m/s). All of the data is taken at 255K. Given this impurity concentration we can estimate the impurity band width to be of order $0.2 eV (2000 K)$. Since most measurements are done at low temperatures, we are always in a regime where $T << |I m \Sigma|$. In this regime, the scattering rate is very weakly dependent on gate voltage up to chemical potentials of order $0.06 eV$.

\subsubsection{Electrical Conductivity}.

The longitudinal conductivity has a plateau around zero gate voltage crossing over to a linear dependence at higher gate voltages. This is consistent with a small impurity bandwidth beyond which the scattering rate is inversely proportional to the energy. Since both the slopes and the crossover scale is determined by the same parameter, the lack agreement is a clear evidence for the departure from the unitary scattering dominated scattering theory.

The observed asymmetric data is specific to the device studied here. The behavior at high carrier densities is consistent with a small impurity concentration of about $1.7 \times 10^{-4}$, but the value at the node requires an impurity concentration that is an order of magnitude smaller. Since the measurements were done using two probes, we have subtracted out the contact resistance. This is accomplished by realizing that the large gate voltage conductivity is linear which extrapolates to zero for graphene. Note that the contact resistance does not affect the thermopower and Nernst measurements, but does affect the Hall resistance.

\begin{figure}
\begin{center}\label{elec-con}
\includegraphics[width= \columnwidth]{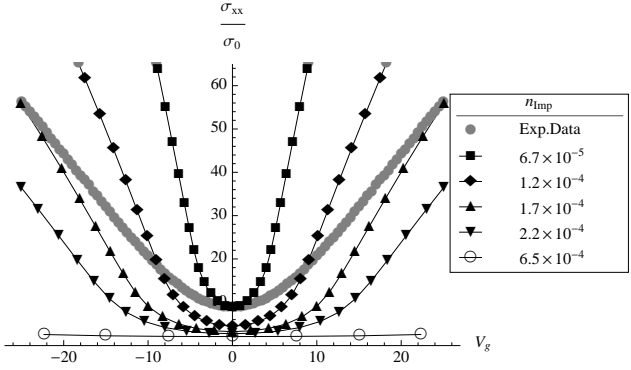}
\caption {Comparison of calculation and experimental data of longitudinal conductivity as a function of gate voltage. The observed conductivity is much larger than those predicted for an impurity concentration of unitary scatterers of $2.2  \times 10^{-4}$ and $6.5  \times 10^{-4}$. An impurity concentration closer to $1.7  \times 10^{-4}$ is consistent with the data.}
\end{center}
\end{figure}

\subsubsection{Hall resistance}

\begin{figure}
\begin{center}
\includegraphics[width= \columnwidth]{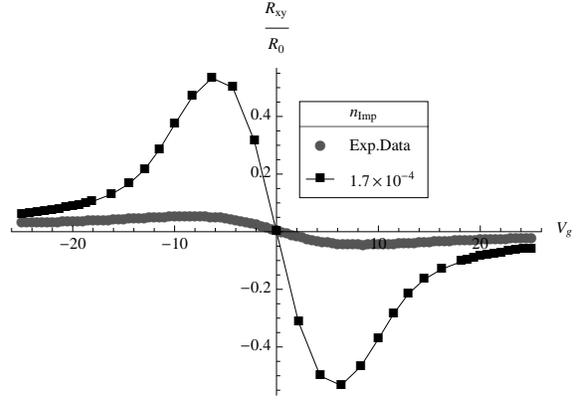}
\caption {The calculated Hall coefficient and experimental data is plotted in units of $R_{0}=\frac {\pi h}{|B|e^2}$ as a function of gate voltage. The data is best fit for impurity concentrations of $1.7  \times 10^{-4}$ for large gate voltages. The overall features are sensitive to the impurity concentration of unitary scatterers as can be seen by the predicted behavior for n=$1.7  \times 10^{-4}$.}
\label{hall-coeff}
\end{center}
\end{figure}

The fit to the observed hall coefficient is shown in fig.\ref{hall-coeff}. The hall varies linearly with gate voltage for small carrier densities and falls off as $1/\mu^{2}$ beyond a scale set by the impurity bandwidth. As mentioned perviously, the two probe measurement of the Hall data is not reliable, and the fact that no agreement with the data is achieved beyond qualitative dependence is expected.

\subsubsection{Thermopower}

\begin{figure}
\begin{center}
\includegraphics[width= \columnwidth] {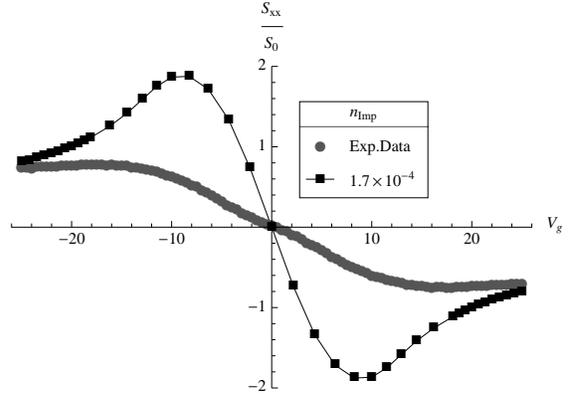}
\caption {The calculated $S_{xx}$ and experimental data plotted in units of $S_{0}=\frac{k_{B}}{e}$ as a function of gate voltage.}
\label{expt-thermopower}
\end{center}
\end{figure}

The measured thermopower in graphene is a linear function of gate voltage for small carrier densities (see fig.\ref{expt-thermopower}). The slope is proportional to $\tau^{2}$. By fitting our numerical solution to the data, we find that the impurity concentration of $6.5 \times 10^{-5}$ (not shown in the figure) can account for the slope of the thermopower. We plot the dependence of thermopower for an impurity concentration of $1.7 \times 10^{-4}$ which agrees with the asymptotic dependence at large carrier density. A single parameter fit for the entire range is not possible.

\subsubsection{Nernst Signal}

\begin{figure}
\begin{center}
\includegraphics[width= \columnwidth] {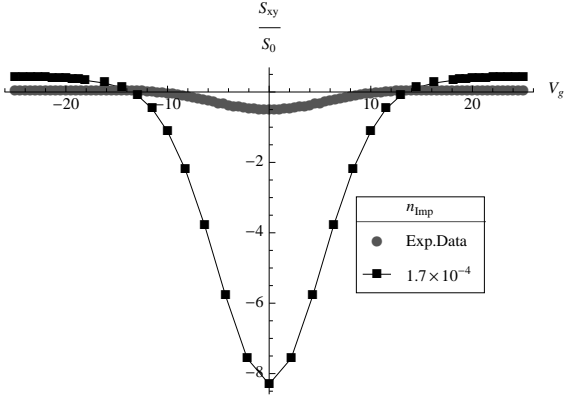}
\caption {Experimental data and calculated $S_{xy}$ plotted in units of $S_{0}=\frac{k_{B}}{e}$ as a function of gate voltage.}
\label{expt-nernst}
\end{center}
\end{figure}

The Nernst signal in graphene is shown in fig.\ref{expt-nernst}. It is negative at zero gate voltage changing sign for large carrier densities. The peak value is large and about $ 50 \mu V/K T$. Theoretically it is proportional to $\tau^{3}$ and is predicted to change sign as a function of chemical potential and gate voltage. The lack of a single parameter fit, as in the case of the thermopower, is obtained for the Nernst as well. The value at the node can be fit with a smaller value of the impurity concentration, but the cross over to the asymptotic behavior is not captured.

The above analysis suggests that a single impurity concentration of unitary scatterers cannot reproduce all the observed data. While no single impurity density fits the conductivity, two different values of impurity concentration, differing by an order of magnitude, are needed to fit the values near the node and the asymptotic behavior at large carrier densities respectively.

\subsection{Coulomb Scatterers}

It is clear from the analysis in the previous section that unitary scatterers fail to accurately reproduce conductivity data. Coulomb scatterers have been shown to reproduce conductivity data at large carrier densities. The fits to the data with Coulomb scatterers is presented in this section. One caveat to note in these fits is that the experimental data for Hall and Nernst are outside the regime of validity of our theoretical calculations in the hydrodynamic limit. In particular the scattering length is much longer than the cyclotron frequency.

\subsubsection{Electrical Conductivity}

\begin{figure}
\begin{center}
\includegraphics[width= \columnwidth]{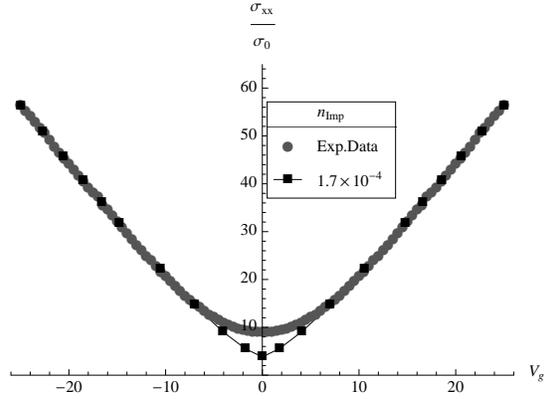}
\caption {The conductivity data and calculations for a charge impurity concentration of Coulomb scatterers of $n_{c}=1.7\times 10^{-4}$. Coulomb scatterers provide excellent quantitative and qualitative agreement in the entire range of carrier densities measured.}
\label{cs}
\end{center}
\end{figure}

The conductivity data can be reproduced over the entire range from low to high carrier concentration (see fig.\ref{cs}). The impurity concentration required is $n_{c}=1.7 \times 10^{-4}$. The finite conductivity at the node is a result of the self consistent treatment of the impurity potential. Even for weak potentials, the induced impurity states provide finite conductivity and screening at the node. The agreement with data suggests that Coulomb  and not unitary scatterers are the predominant source of scattering in these systems.

\subsubsection{Hall Resistance}

\begin{figure}
\begin{center}
\includegraphics[width= \columnwidth]{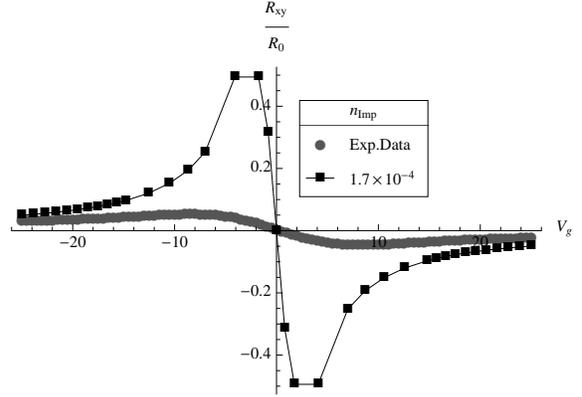}
\caption {Hall resistivity data fit to calculations for a charge impurity concentration of $n_{c}=1.7 \times10^{-4}$, and is plotted in units of $R_{0}=\frac {\pi h}{|B|e^2}$ as a function of gate voltage}
\label{crxy}
\end{center}
\end{figure}

The fit to the observed hall coefficient is shown in fig.\ref {crxy}. The hall varies linearly with gate voltage for small carrier densities and falls off as $1/\mu^{2}$ beyond a scale set by the impurity bandwidth. The lack of the agreement is apparent, but the same drawback of two terminal measurement precludes any quantitative conclusions from being drawn.

\subsubsection{Thermopower}

\begin{figure}
\begin{center}
\includegraphics[width= \columnwidth]{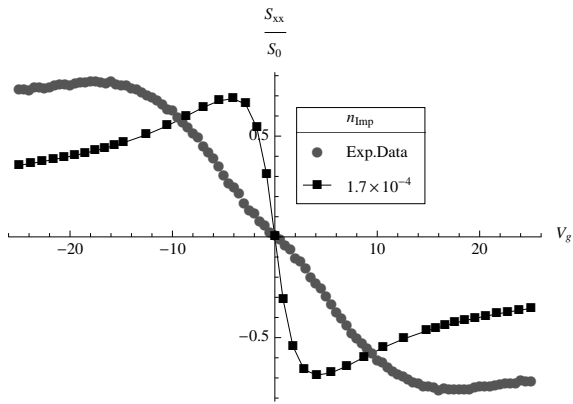}
\caption {Thermpower data fit to calculations for a charge impurity concentration of $n_{c}=1.7 \times10^{-4}$.}
\label{csxx}
\end{center}
\end{figure}

The overall agreement between Thermopower data and the predictions from Coulomb scattering dominated transport is significantly better than that for unitary scatterers(see fig.\ref{csxx}). The peak's position and slope are off by a factor of $O(1)$ ($\sim 2$). Since thermopower is sensitive to higher derivatives of the scattering rate with respect to energy as compared to electrical conductivity, this reflects the difference in the dependence of the imaginary part of the self energy of unitary and charge scattering potentials (see fig.\ref{selfenergy} and fig.\ref{chselfenergy}). In particular, the self energy in the former varies by a factor of $\sim 4$ while the latter changes by a factor of $\sim 2$ for a change in gate voltage from $0$ to $0.1$ eV. The weaker dependence of Coulomb scatterers provides a much better fit to the data.

\subsubsection{Nernst Signal}

\begin{figure}
\begin{center}
\includegraphics[width= \columnwidth]{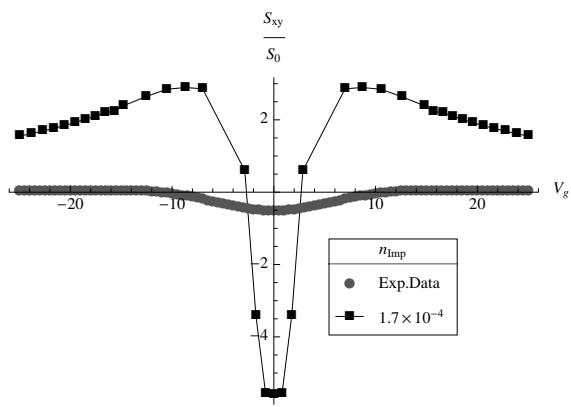}
\caption {Nernst signal data fit to calculations for a charge impurity concentration of $n_{c}=1.7 \times10^{-4}$.}
\label{csxy}
\end{center}
\end{figure}

The observed peak in the Nernst signal at the node is smaller by an order of magnitude than that expected from charge scatterers (see fig.\ref{csxy}). This result is consistent with the observation that charge scatterers are not sufficient to accurately reproduce thermopower data, since both depend on the variation of the scattering rate as a function of energy. Furthermore, the crossover to the asymptotic behavior is not captured by the Coulomb scattering phenomenology. In particular, the positive peaks are not seen in the data.

A simplifying assumption made in these calculations is that the self energy is independent of momentum. Including the momentum dependence similar agreement with conductivity \cite{PhysRevB.77.125409} and Hall coefficient \cite{PhysRevB.80.155423} has been obtained for finite range scatterers. The inclusion of momentum dependence also yields better agreement with thermopower data \cite{PhysRevB.80.165423} while the discrepancy with Nernst data cannot be resolved with this generalization \cite{PhysRevB.81.155457}. 

\section{Conclusion and Acknowledgments}

In this comprehensive study of the transport coefficients on graphene, we show that the theories assuming a single scattering mechanism, either charged Coulomb or unitary scattering, are incapable of accounting for all the observed data. A number of studies that focus only on electrical conductivity emphasize that one or the other of these two mechanisms as the source of the minimum conductivity and the linear in gate voltage dependence. Since the only fitting parameter is the impurity density, our calculations show that charged coulomb rather than unitary scatterers fit the observed conductivity data. In particular if one chooses to fit the minimum conductivity using unitary scatterers, the slope at high gate voltages is overestimated, while a fit to slope at high gate voltages underestimates the minimum conductivity. Given this observation one would conclude that unitary scatterers are not the dominant source of scattering in graphene.

To check the validity of the conclusion based on electrical conductivity, we fit the data for hall resistance, thermopower and Nernst all obtained on the same sample. This is important because fitting data on different samples require choosing different impurity concentrations which render a quantitative comparison invalid. Moreover, the dominant scattering mechanisms in different samples need not be the same. Analytic results in asymptotic regimes are provided for all transport coefficients, and numerical calculations are used to fit the data over a wide range of gate voltage.

The most striking disagreement is in the Nernst Data near the node. Both types of impurities overestimate the value by an order of magnitude. The Nernst Data asymptotic values are recovered for the unitary scatterers while the Coulomb scatterers over estimate these values by two orders of magnitude. In other words in the regime where the Coulomb scatterers fit conductivity data, the Nernst is significantly overestimated. The inability of a single scattering mechanism to get both the behavior near the node and at large carrier densities for all observed transports coefficients is the principal conclusion of our work.

Finally, a few comments on the validity of the theoretical calculations are merited. We have used the SCBA formalism to compute the transport characteristics. While the approximation fails in the vicinity of the node due to interference effects, the regime of validity is large as seen in the agreement of the density of states determined within this approximation and exact numerical methods [see \onlinecite{peresrmp} and references therein]. A second caveat is near the node the effect of the magnetic fields cannot be captured within the semiclassical treatment presented. The Landau level splitting becomes larger than the Fermi energy for a field as low as 1T field when doping level is of the order $10^{-4}$. For higher chemical potentials, the transport calculations are reliable. However, within a small window near the node both the SCBA and the magnetic field effects need to accounted for.

The authors wish to acknowledge Jing Shi and Peng Wei for useful discussions. This research has been supported by University of California at Riverside.

\appendix

\section{Electrical conductivity tensor in a Homogeneous Magnetic Field}

\noindent In this appendix, we define the electrical current density in the presence of a homogeneous magnetic field and derive conductivity tensor in the presence of a finite range disorder. The method developed by Khodas and Finkel'stein to calculate the Hall coefficient is used \cite{PhysRevB.68.155114}. The conductivity is calculated within linear response. The electrical current density is\cite{PhysRevB.68.155114}
 
 \begin{widetext}
 \begin{eqnarray}
 j_{e,\alpha}(r_{i},t)&=&  \lim_{ \begin{subarray}{1} r_{i}' \to r_{i} \\ t' \to t \end{subarray}} \frac{e}{2m}[(-i\overrightarrow{\nabla}_{r_{i}'}^{\alpha} -e\overrightarrow{A}_ {r_{i}'}^{\alpha}) - (-i\overrightarrow{\nabla}_{r_{i}}^{\alpha} +e\overrightarrow{A}_ {r_{i}}^{\alpha})  ]\Psi^{\dagger}(r_{i},t)\Psi(r_{i}',t')\
\end{eqnarray}
\end{widetext}
 
\noindent The electrical current-current correlation function is 
 
 \begin{widetext}
 \begin{eqnarray}\label{e-e}
 \Pi_ {\alpha,\beta} ^{e,e}(r_{i},r_{f};\tau) &=& <T_{\tau}(\lim_ { \begin{subarray}{1} r_{i}' \to r_{i} \\ \tau' \to \tau \end{subarray}} \frac{e}{2m}[(-i\overrightarrow{\nabla}_{r_{i}'}^{\alpha} -e\overrightarrow{A}_ {r_{i}'}^{\alpha}) - (-i\overrightarrow{\nabla}_{r_{i}}^{\alpha} +e\overrightarrow{A}_ {r_{i}}^{\alpha})  ]\Psi^{\dagger}(r_{i},\tau)\Psi(r_{i}',\tau'))\nonumber\\
 & &(\lim_{ r_ {f}' \to r_ {f}} \frac{e}{2m}[(i\overrightarrow{\nabla}_{r_ {f}'}^{\beta} -e\overrightarrow{A}_ {r_ {f}'}^{\beta}) - (i\overrightarrow{\nabla}_{r_ {f}}^{\beta} +e\overrightarrow{A}_ {r_ {f}}^{\beta})  ]\Psi^{\dagger}(r_ {f}',0)\Psi(r_ {f},0)) >\
 \end{eqnarray}
\end{widetext}

\noindent In the presence of a magnetic field the Greens function is defined as $G(r_{1},r_{2},\tau)= exp(\frac{\imath e}{c}\Phi (r_{1},r_{2}))\widetilde{G}_{r_{1}-r_{2}}(\tau)$, where $\widetilde{G}_{r_{1}-r_{2}}(\tau)$ is the gauge invariant Greens function, and the exponential factor accounts for the phase acquired by the particle along the path $\overrightarrow{r}_{i} \rightarrow \overrightarrow{r}_{f}$\cite {PhysRevB.68.155114}. Choosing the vector potential as $\overrightarrow{A} = \frac{1}{2} \overrightarrow{B}\times\overrightarrow{r}$ and using the form of the phase factor of the exponential of the Green's function $\Phi (r_{1},r_{2})=\int_{r_{1}}^{r_{2}} \overrightarrow {A} (\overrightarrow{r}) d \overrightarrow {r}$, eqn.\ref{e-e} can be written as

\begin{widetext}
\begin{eqnarray}
 \Pi_ {\alpha,\beta} ^{e,e}(r_{i},r_{f};\tau) &=& \Pi_ {\alpha,\beta} ^{e,e,1}(r_{i},r_{f};\tau) + \Pi_ {\alpha,\beta}^{e,e,2} (r_{i},r_{f};\tau)  + \Pi_ {\alpha,\beta} ^{e,e,3}(r_{i},r_{f};\tau)  + \Pi_ {\alpha,\beta} ^{e,e,4}(r_{i},r_{f};\tau) \\
\Pi_ {\alpha,\beta} ^{e,e,1}(r_{i},r_{f};\tau) &=&\frac{-e^{2}} {4m^2}[(-i\overrightarrow{\nabla}_{r_{i}} -\frac{e\overrightarrow{B}}{2c}\times(\overrightarrow{r_{i}} - \overrightarrow{r_{f}}) )^ {\alpha} (i\overrightarrow{\nabla}_{r_{f}} +\frac{e\overrightarrow{B}}{2c}\times(\overrightarrow{r_{i}} - \overrightarrow{r_{f}}))^{\beta}\widetilde{G}_{r_{i}-r_{f}}(\tau)](\widetilde{G}_{r_{f}-r_{i}}(-\tau)) \nonumber\\
\Pi_ {\alpha,\beta} ^{e,e,2}(r_{i},r_{f};\tau) &=& \frac {-e^{2}} {4m^2}(\widetilde{G}_{r_{i}-r_{f}}(\tau))[(-i\overrightarrow{\nabla}_{r_{i}} +\frac{e\overrightarrow{B}}{2c}\times(\overrightarrow{r_{i}} - \overrightarrow{r_{f}}))^{\alpha} (i\overrightarrow{\nabla}_{r_{f}} - \frac{e\overrightarrow{B}}{2c}\times(\overrightarrow{r_{i}} - \overrightarrow{r_{f}}))^ {\beta}\widetilde{G}_{r_{f}-r_{i}}(-\tau) \nonumber\\
\Pi_ {\alpha,\beta} ^{e,e,3}(r_{i},r_{f};\tau) &=& \frac {e^{2}} {4m^2}[(-i\overrightarrow{\nabla}_{r_{i}} -\frac{e\overrightarrow{B}}{2c}\times(\overrightarrow{r_{i}} - \overrightarrow{r_{f}}))^ {\alpha} \widetilde{G}_{r_{i}-r_{f}}(\tau)][(i\overrightarrow{\nabla}_{r_{f}} -\frac{e\overrightarrow{B}}{2c}\times(\overrightarrow{r_{i}} - \overrightarrow{r_{f}}))^ {\beta} \widetilde{G}_{r_{f}-r_{i}}(-\tau)] \nonumber\\
\Pi_ {\alpha,\beta} ^{e,e,4}(r_{i},r_{f};\tau) &=& \frac {e^{2}}{4m^2}[(-i\overrightarrow{\nabla}_{r_{i}} +\frac{e\overrightarrow{B}}{2c}\times(\overrightarrow{r_{i}} - \overrightarrow{r_{f}}))^ {\alpha} \widetilde{G}_{r_{f}-r_{i}}(-\tau)][(i\overrightarrow{\nabla}_{r_{f}} +\frac{e\overrightarrow{B}}{2c}\times(\overrightarrow{r_{i}} - \overrightarrow{r_{f}}))^ {\beta}(\widetilde{G}_{r_{i}-r_{f}}(\tau))] \nonumber\\
\end{eqnarray}
\end{widetext}

\noindent Letting $\overrightarrow{R}=\overrightarrow{r}_{i}-\overrightarrow{r}_{f}$, taking a fourier transform of each of these correlation functions and labeling $\overrightarrow{K}=\overrightarrow{p}+\overrightarrow{q}$ we get

\begin{widetext}
\begin{eqnarray}
 \Pi_ {\alpha,\beta} ^{e,e}(\overrightarrow{q};i\Omega) &=& \Pi_ {\alpha,\beta} ^{e,e,1} (\overrightarrow{q};i\Omega) + \Pi_ {\alpha,\beta}^{e,e,2} (\overrightarrow{q};i\Omega)  + \Pi_ {\alpha,\beta} ^{e,e,3} (\overrightarrow{q};i\Omega)  + \Pi_ {\alpha,\beta} ^{e,e,4} (\overrightarrow{q};i\Omega) \\
 \Pi_ {\alpha,\beta} ^{e,e,1} (\overrightarrow{q};i\Omega) &=&\frac{e^{2}} {4m^2 \beta}\sum_{\overrightarrow{p},iw_{n}}[(\overrightarrow{K} +i\frac{e\overrightarrow{B}}{2c}\times \overrightarrow{\nabla}_{\overrightarrow{K}} )^ {\alpha} (\overrightarrow{K} -i\frac{e\overrightarrow{B}}{2c}\times\overrightarrow{\nabla}_{\overrightarrow{K}})^{\beta}\widetilde{G}_{\overrightarrow{K}}(iw_{n} + i\Omega)](\widetilde{G}_{\overrightarrow{p}}(iw_{n})) \nonumber\\
\Pi_ {\alpha,\beta} ^{e,e,2} (\overrightarrow{q};i\Omega) &=& \frac{e^{2}} {4m^2 \beta} \sum_{\overrightarrow{p},iw_{n}}\widetilde{G}_{\overrightarrow{K}}(iw_{n}+i \Omega)[(-\overrightarrow{p} +i\frac{e\overrightarrow{B}}{2c}\times \overrightarrow{\nabla}_{\overrightarrow {p}} ) ^{\alpha} (-\overrightarrow{p} -i\frac{e\overrightarrow{B}}{2c}\times\overrightarrow{\nabla}_{\overrightarrow {p}}) ^ {\beta}\widetilde{G}_{\overrightarrow{p}}(iw_{n})] \nonumber\\
 \Pi_ {\alpha,\beta} ^{e,e,3} (\overrightarrow{q};i\Omega) &=& \frac{-e^{2}} {4m^2 \beta}\sum_{\overrightarrow{p},iw_{n}}[(\overrightarrow{K} +i\frac{e\overrightarrow{B}}{2c}\times \overrightarrow{\nabla}_{\overrightarrow{K}} ) ^ {\alpha} \widetilde{G}_{\overrightarrow{K}}(iw_{n}+i \Omega)][(-\overrightarrow{p}-i\frac{e\overrightarrow{B}}{2c}\times\overrightarrow{\nabla}_{\overrightarrow {p}}) ^ {\beta} (\widetilde{G}_{\overrightarrow{p}}(iw_{n}))] \nonumber\\
 \Pi_ {\alpha,\beta} ^{e,e,4} (\overrightarrow{q};i\Omega) &=& \frac{-e^{2}} {4m^2 \beta}\sum_{\overrightarrow{p},iw_{n}}[(-\overrightarrow{p} +i\frac{e\overrightarrow{B}}{2c}\times \overrightarrow{\nabla}_{\overrightarrow {p}} )^ {\alpha} \widetilde{G}_{\overrightarrow{p}}(iw_{n})][(\overrightarrow{K} -i\frac{e\overrightarrow{B}}{2c}\times\overrightarrow{\nabla}_{\overrightarrow{K}}) ^ {\beta}(\widetilde{G}_{\overrightarrow{K}}(iw_{n}+i \Omega))] \nonumber\\
\end{eqnarray}
\end{widetext}

\noindent To first order in magnetic field (assumed to be in the positive z direction) the diagonal and off diagonal terms are

\begin{widetext}
\begin{eqnarray}
 \Pi_ {x,x} ^{e,e} (\overrightarrow{q}\to0;i\Omega) &=&\frac{e^{2}} {\beta}\sum_{\overrightarrow{p},iw_{n}} \{(\overrightarrow{v}_{\overrightarrow{p},x}^{2}) \widetilde{G}_ {\overrightarrow{p}}(iw_{n} + i\Omega) \widetilde{G}_ {\overrightarrow{p}} (iw_{n}) \}\\
 \Pi_ {x,y} ^{e,e} (\overrightarrow{q}\to0;i\Omega) &=&\frac{-e^{3}i| \overrightarrow {B}|}{4c\beta m}\sum_{\overrightarrow{p},iw_{n}} (\overrightarrow{v}_{\overrightarrow{p},x})\{\frac{\partial \widetilde{G}_ {\overrightarrow{p}}(iw_{n} + i\Omega)}{\partial p_{x}} \widetilde{G}_ {\overrightarrow{p}}(iw_{n}) - \widetilde{G}_ {\overrightarrow{p}}(iw_{n} + i\Omega) \frac{\partial \widetilde{G}_ {\overrightarrow{p}}(iw_{n})}{\partial p_{x}}\} \nonumber\\
 &+&\frac{-e^{3}i| \overrightarrow {B}|} {4c\beta m}\sum_{\overrightarrow{p},iw_{n}} (\overrightarrow{v}_{\overrightarrow{p},y})\{\frac{\partial \widetilde{G}_ {\overrightarrow{p}}(iw_{n} + i\Omega)}{\partial p_{y}} \widetilde{G}_ {\overrightarrow{p}}(iw_{n}) -\widetilde{G}_ {\overrightarrow{p}}(iw_{n} + i\Omega) \frac{\partial \widetilde{G}_ {\overrightarrow{p}}(iw_{n})}{\partial p_{y}}\}\
\end{eqnarray}
\end{widetext}

\noindent The current-current correlation tensor and impurity self averaging Green's function\cite{PhysRevB.73.125411} of graphene which has two bands is

\begin{widetext}
\begin{eqnarray}
 \Pi_ {x,x} ^{e,e} (\overrightarrow{q}\to0;i\Omega) &=&\frac{N_{v}e^{2}} {\beta}\sum_{\overrightarrow{p},iw_{n}} \{(\overrightarrow{v}_{\overrightarrow{p},x}^{2}) \widetilde{G}_ {AA,\overrightarrow{p}}(iw_{n} + i\Omega) \widetilde{G}_ {AA,\overrightarrow{p}} (iw_{n}) \}\\
 \Pi_ {x,y} ^{e,e} (\overrightarrow{q}\to0;i\Omega) &=&\frac{-N_{v}e^{3}i| \overrightarrow {B}|}{c\beta m}\sum_{\overrightarrow{p},iw_{n}} (\overrightarrow{v}_{\overrightarrow{p}})^{2} \widetilde{G}_ {AA,\overrightarrow{p}}(iw_{n} + i\Omega)\widetilde{G}_ {C,\overrightarrow{p}}(iw_{n} + i\Omega) \widetilde{G}_ {AA,\overrightarrow{p}}(iw_{n}) \nonumber\\
 &+&\frac{Ne^{3}i| \overrightarrow {B}|}{c\beta m}\sum_{\overrightarrow{p},iw_{n}} (\overrightarrow{v}_{\overrightarrow{p}})^{2} \widetilde{G}_ {AA,\overrightarrow{p}}(iw_{n} + i\Omega) \widetilde{G}_ {AA,\overrightarrow{p}}(iw_{n}) \widetilde{G}_ {C,\overrightarrow{p}}(iw_{n})\\
G_{(AA,BB)}(\overrightarrow{K},\mathrm{i}\varepsilon) &=& \frac{(\mathrm{i}\varepsilon-\Sigma{(\mathrm{i}\varepsilon)})}{(\mathrm{i}\varepsilon-\Sigma{(\mathrm{i}\varepsilon)}) ^2-|\phi(\overrightarrow{K})|^{2}}\\
G_{C}(\overrightarrow{K},\mathrm{i}\varepsilon) &=&\frac{|\phi(\overrightarrow{K})|} {(\mathrm{i}\varepsilon-\Sigma{(\mathrm{i}\varepsilon)})^2-|\phi(\overrightarrow{K})|^{2}}\
\end{eqnarray}
\end{widetext}

\noindent Performing the sum over frequency and taking the real part of the electrical conductivity tensor we get

\begin{widetext}
\begin{eqnarray}
	\sigma_{xx} &=& \frac{N_{v}e^{2}}{V}\sum_{\overrightarrow{p}}\int \frac{d\varepsilon}{\pi} (\frac{n_{F}(\varepsilon)-n_{F} (\varepsilon+\Omega)}{\Omega}) \overrightarrow{v}_{\overrightarrow{p},x}^{2} ImG_{AA,p} (\varepsilon) ImG_{AA,p}(\varepsilon+\Omega)\\
	\sigma_{xy} &=& \frac{-N_{v}e^{3}| \overrightarrow {B}|v_{F}^{2}}{c\Omega V}\sum_{\overrightarrow{p}}\int \frac{d\varepsilon}{\pi m} (n_{F}(\varepsilon)(\Sigma^{(1,a)}_{xy}-\Sigma^{(2,a)}_{xy})- n_{F}(\varepsilon+\Omega) (\Sigma^{(1,b)}_{xy}-\Sigma^{(2,b)}_{xy}))\\
	\Sigma^ {(1,a)}_{xy} &=& Im\widetilde{G}_ {AA,\overrightarrow{p}}(\varepsilon)\{ Re\widetilde{G}_ {AA,\overrightarrow{p}}(\varepsilon+\Omega)Re\widetilde{G}_ {C,\overrightarrow{p}}(\varepsilon+\Omega)-Im\widetilde{G}_ {AA,\overrightarrow{p}}(\varepsilon+\Omega)Im\widetilde{G}_ {C,\overrightarrow{p}}(\varepsilon+\Omega)\}\nonumber\\
	\Sigma^ {(2,a)}_{xy} &=& Re\widetilde{G}_ {AA,\overrightarrow{p}}(\varepsilon+\Omega)\{ Im\widetilde{G}_ {AA,\overrightarrow{p}}(\varepsilon)Re\widetilde{G}_ {C,\overrightarrow{p}}(\varepsilon)+Re\widetilde{G}_ {AA,\overrightarrow{p}}(\varepsilon)Im\widetilde{G}_ {C,\overrightarrow{p}}(\varepsilon)\}\nonumber\\
	\Sigma^{(1,b)}_{xy} &=& Im\widetilde{G}_ {AA,\overrightarrow{p}} (\varepsilon+\Omega)\{ Re\widetilde{G}_ {AA,\overrightarrow{p}} (\varepsilon) Re\widetilde{G}_ {C,\overrightarrow{p}} (\varepsilon)-Im\widetilde{G}_ {AA,\overrightarrow{p}} (\varepsilon) Im\widetilde{G}_ {C,\overrightarrow{p}} (\varepsilon)\}\nonumber\\
	\Sigma^ {(2,b)}_{xy} &=& Re\widetilde{G}_ {AA,\overrightarrow{p}} (\varepsilon)\{ Im\widetilde{G}_ {AA,\overrightarrow{p}} (\varepsilon+\Omega) Re\widetilde{G}_ {C,\overrightarrow{p}} (\varepsilon+\Omega) +Re\widetilde{G}_ {AA,\overrightarrow{p}} (\varepsilon+\Omega) Im\widetilde{G}_ {C,\overrightarrow{p}} (\varepsilon+\Omega)\}\nonumber\\
\end{eqnarray}
\end{widetext}

\noindent Integrating over momentum and taking the limit of an ideal Dirac spectrum the diagonal conductivity in the dc limit is

\begin{widetext}
\begin{equation}
	\sigma_{xx}^{DC} = \frac{N_{s}N_{v}e^2}{4\pi h} \int d\varepsilon \frac{\partial n_{F}}{\partial \mu}(1+\frac{A^2+B^2}{AB} \arctan{\frac {A}{B}})
\end{equation}
\end{widetext}

\noindent where A and B are defined as $\varepsilon - Re\Sigma(\varepsilon)$ and $Im\Sigma(\varepsilon)$.
The hall term can be simplified by letting $\Omega \rightarrow 0$ and using $(ReG_{BB}^{2}+ImG_{BB}^{2}) ImG_{C}= Im(|G_{BB}|^{2}G_{C})$, $(ReG_{BB}^{2}-ImG_{BB}^{2}) \frac{\partial ImG_{C}}{\partial \varepsilon} + 2ReG_{BB} ImG_{BB} \frac{\partial ReG_{C}}{\partial \varepsilon})= Im(G_{BB}^{2}\frac{\partial G_{C}} {\partial \varepsilon})$. 

\begin{widetext}
\begin{eqnarray}
	\sigma_{xy}^{DC} &=& \frac{N_{v}e^{3}| \overrightarrow {B}|v_{F}^{3}}{cV}\sum_{\overrightarrow{p}}\int \frac{d\varepsilon}{ \pi | \overrightarrow{p}|}\{ \frac{\partial n_{F}(\varepsilon)}{\partial \mu}(|\widetilde{G}_ {AA,\overrightarrow{p}} (\varepsilon)|^{2} Im \widetilde{G}_ {C,\overrightarrow{p}} (\varepsilon) ) - n_{F}(\varepsilon) Im(\widetilde{G}_ {AA,\overrightarrow{p}}^ {2} (\varepsilon)\frac{\partial \widetilde{G}_ {C,\overrightarrow{p}} (\varepsilon)}{\partial \varepsilon} )\}\
\end{eqnarray}
\end{widetext}

\noindent Calculating the angular integral first the hall term reduces to

\begin{widetext}
\begin{eqnarray}
\sigma_{xy}^{DC} &=& \frac{N_{s}N_{v}e^{3}| \overrightarrow {B}|v_{F}^{3}}{2c\pi} \int dp \int \frac{d\varepsilon}{\pi}\{ \frac{\partial n_{F}(\varepsilon)}{\partial \mu}(|\widetilde{G}_ {AA,\overrightarrow{p}} (\varepsilon)|^{2} Im \widetilde{G}_ {C,\overrightarrow{p}} (\varepsilon) ) - n_{F}(\varepsilon) Im(\widetilde{G}_ {AA,\overrightarrow{p}}^ {2} (\varepsilon)\frac{\partial G_{C}} {\partial \varepsilon})\}\
\end{eqnarray}
 
 \noindent where the integrals of the green's functions are
 
\begin{eqnarray}
\int d x(|\widetilde{G}_ {AA,x} (\varepsilon)|^{2} Im \widetilde{G}_ {C,x} (\varepsilon) ) &=& \frac{-1}{8AB}(\frac{B^{2}-A^{2}}{B^{2}+A^{2}} - \frac{B^{2}+A^{2}}{2AB} \arctan{\frac{2AB}{B^{2}-A^{2}}})\\
\int d x Im (\widetilde{G}_ {AA,x} (\varepsilon)^{2}\frac{\partial \widetilde{G}_ {C,x} (\varepsilon)}{\partial \varepsilon} ) &=& \frac{-1}{3}\frac{\partial}{\partial \varepsilon}(\frac{AB}{(A^{2}+B^{2})^{2}})\
\end{eqnarray}
\end{widetext}

\noindent Integrating by parts,

\begin{widetext}
\begin{eqnarray}
\sigma_{xy}^{DC} &=& \frac{-N_{s}N_{v}e^{3}| \overrightarrow {B}|v_{F}^{2}}{2c\pi}\int \frac{d\varepsilon}{\pi} \frac{\partial n_{F}(\varepsilon)}{\partial \mu} \{ \frac{1}{8AB}(\frac{B^{2}-A^{2}}{B^{2}+A^{2}} - \frac{B^{2}+A^{2}}{2AB} \arctan{\frac{2AB}{B^{2}-A^{2}}}) - \frac{AB}{3(A^{2}+B^{2})^{2}}\}\
\end{eqnarray}
\end{widetext}

 \section{Thermoelectric conductivity tensors in a Homogeneous Magnetic Field }

\noindent In this appendix, we define the electrical and energy current density in the presence of a homogeneous magnetic field. We follow the same approach as we did for longitudinal conductivity in the previous section. The currents are\cite {PhysRevB.68.155114,PhysRevB.67.115131}

 \begin{widetext}
 \begin{eqnarray}
 j_{e,\alpha}(r_{i},t)&=&  \lim_{ \begin{subarray}{1} r_{i}' \to r_{i} \\ t' \to t \end{subarray}} \frac{e}{2m}[(-i\overrightarrow{\nabla}_{r_{i}'}^{\alpha} -e\overrightarrow{A}_ {r_{i}'}^{\alpha}) - (-i\overrightarrow{\nabla}_{r_{i}}^{\alpha} +e\overrightarrow{A}_ {r_{i}}^{\alpha})  ]\Psi^{\dagger}(r_{i},t)\Psi(r_{i}',t')\\
j_{E,\alpha}(r_{i},t)&=&  \lim_ { \begin{subarray}{1} r_{i}' \to r_{i} \\ t' \to t \end{subarray}} \frac{1}{2m}[(-i\overrightarrow{\nabla}_{r_{i}'}^{\alpha} -e\overrightarrow{A}_ {r_{i}'}^{\alpha})\frac{\partial}{\partial t} + (-i\overrightarrow{\nabla}_{r_{i}}^{\alpha} +e\overrightarrow{A}_ {r_{i}}^{\alpha})\frac{\partial}{\partial t'}  ]\Psi^{\dagger}(r_{i},t)\Psi(r_{i}',t')\
\end{eqnarray}
 
 \noindent Letting $t \to i\tau$ the energy current becomes
 
 \begin{eqnarray}
j_{E,\alpha}(r_{i},t)&=&  \lim_ { \begin{subarray}{1} r_{i}' \to r_{i} \\ \tau' \to \tau \end{subarray}} \frac{i}{2m}[(-i\overrightarrow{\nabla}_{r_{i}'}^{\alpha} -e\overrightarrow{A}_ {r_{i}'}^{\alpha})\frac{\partial}{\partial \tau} + (-i\overrightarrow{\nabla}_{r_{i}}^{\alpha} +e\overrightarrow{A}_ {r_{i}}^{\alpha})\frac{\partial}{\partial \tau'}  ]\Psi^{\dagger}(r_{i},\tau)\Psi(r_{i}',\tau')\
\end{eqnarray}
 
\noindent The correlation function that determines the thermoelectric response is

 \begin{eqnarray}\label{eh}
 \Pi_ {\alpha,\beta} ^{E,e}(r_{i},r_{f};\tau) &=& <T_{\tau}(\lim_ { \begin{subarray}{1} r_{i}' \to r_{i} \\ \tau' \to \tau \end{subarray}} \frac{i}{2m}[(-i\overrightarrow{\nabla}_{r_{i}'}^{\alpha} -e\overrightarrow{A}_ {r_{i}'}^{\alpha})\frac{\partial}{\partial \tau} + (-i\overrightarrow{\nabla}_{r_{i}}^{\alpha} +e\overrightarrow{A}_ {r_{i}}^{\alpha})\frac{\partial}{\partial \tau'}  ]\Psi^{\dagger}(r_{i},\tau)\Psi(r_{i}',\tau'))\nonumber\\
 & &(\lim_{ r_ {f}' \to r_ {f}} \frac{e}{2m}[(i\overrightarrow{\nabla}_{r_ {f}'}^{\beta} -e\overrightarrow{A}_ {r_ {f}'}^{\beta}) - (i\overrightarrow{\nabla}_{r_ {f}}^{\beta} +e\overrightarrow{A}_ {r_ {f}}^{\beta})  ]\Psi^{\dagger}(r_ {f}',0)\Psi(r_ {f},0)) >
 \end{eqnarray}
\end{widetext}

\noindent  The greens function $\widetilde{G}_{r_{1}-r_{2}}(\tau)$ is found using the impurity self averaging technique\cite{PhysRevB.73.125411}. Since we have terms that depend on the derivative with respect to $\tau$, we use the equations of motion to determine the greens functions. Using $\frac{\partial}{\partial \tau}G(r_{1},r_{2},\pm\tau) = \pm\delta(\tau)\delta (r_{1}-r_{2}) + < T_{\tau} \frac{\partial}{\partial \tau}\Psi(r_{1},\tau) \Psi^{\dagger}(r_{2},0) > $, eqn.\ref{eh} is

\begin{widetext}
\begin{eqnarray}
 \Pi_ {\alpha,\beta} ^{E,e}(r_{i},r_{f};\tau) &=& \Pi_ {\alpha,\beta} ^{E,e,1}(r_{i},r_{f};\tau) + \Pi_ {\alpha,\beta}^{E,e,2} (r_{i},r_{f};\tau)  + \Pi_ {\alpha,\beta} ^{E,e,3}(r_{i},r_{f};\tau)  + \Pi_ {\alpha,\beta} ^{E,e,4}(r_{i},r_{f};\tau) \\
 \Pi_ {\alpha,\beta} ^{E,e,1}(r_{i},r_{f};\tau) &=&\frac{ei} {4m^2}[(i\overrightarrow{\nabla}_{r_{i}} -\frac{e\overrightarrow{B}}{2c}\times(\overrightarrow{r_{f}} - \overrightarrow{r_{i}}) )^ {\alpha} (i\overrightarrow{\nabla}_{r_{f}} -\frac{e\overrightarrow{B}}{2c}\times(\overrightarrow{r_{f}} - \overrightarrow{r_{i}}))^{\beta}\widetilde{G}_{r_{i}-r_{f}}(\tau)](\frac{\partial \widetilde{G}_{r_{f}-r_{i}}(-\tau)}{\partial \tau}) \nonumber\\
&-&\frac{ei} {4m^2}[(i\overrightarrow{\nabla}_{r_{i}} -\frac{e\overrightarrow{B}}{2c}\times(\overrightarrow{r_{f}} - \overrightarrow{r_{i}}) )^ {\alpha} (i\overrightarrow{\nabla}_{r_{f}} -\frac{e\overrightarrow{B}}{2c}\times(\overrightarrow{r_{f}} - \overrightarrow{r_{i}}))^{\beta}\widetilde{G}_{r_{i}-r_{f}}(\tau)](\delta(-\tau)\delta(r_{f}-r_{i})) \nonumber\\
\Pi_ {\alpha,\beta} ^{E,e,2}(r_{i},r_{f};\tau) &=& \frac {-ei} {4m^2}(\frac{\partial \widetilde{G}_{r_{i}-r_{f}}(\tau)}{\partial \tau})[(i\overrightarrow{\nabla}_{r_{i}} +\frac{e\overrightarrow{B}}{2c}\times(\overrightarrow{r_{f}} - \overrightarrow{r_{i}}))^{\alpha} (i\overrightarrow{\nabla}_{r_{f}} + \frac{e\overrightarrow{B}}{2c}\times(\overrightarrow{r_{f}} - \overrightarrow{r_{i}}))^ {\beta}\widetilde{G}_{r_{f}-r_{i}}(-\tau) \nonumber\\
&+& \frac {-ei} {4m^2}(\delta(\tau)\delta(r_{i}-r_{f}))[(i\overrightarrow{\nabla}_{r_{i}} +\frac{e\overrightarrow{B}}{2c}\times(\overrightarrow{r_{f}} - \overrightarrow{r_{i}}))^{\alpha} (i\overrightarrow{\nabla}_{r_{f}} + \frac{e\overrightarrow{B}}{2c}\times(\overrightarrow{r_{f}} - \overrightarrow{r_{i}}))^ {\beta}\widetilde{G}_{r_{f}-r_{i}}(-\tau) \nonumber\\
 \Pi_ {\alpha,\beta} ^{E,e,3}(r_{i},r_{f};\tau) &=& \frac{-ei} {4m^2}[(i\overrightarrow{\nabla}_{r_{i}} -\frac{e\overrightarrow{B}}{2c}\times(\overrightarrow{r_{f}} - \overrightarrow{r_{i}}))^ {\alpha} \widetilde{G}_{r_{i}-r_{f}}(\tau)][(i\overrightarrow{\nabla}_{r_{f}} +\frac{e\overrightarrow{B}}{2c}\times(\overrightarrow{r_{f}} - \overrightarrow{r_{i}}))^ {\beta} (\frac{\partial \widetilde{G}_{r_{f}-r_{i}}(-\tau)}{\partial \tau})] \nonumber\\
 &+& \frac{ei} {4m^2}[(i\overrightarrow{\nabla}_{r_{i}} -\frac{e\overrightarrow{B}}{2c}\times(\overrightarrow{r_{f}} - \overrightarrow{r_{i}}))^ {\alpha} \widetilde{G}_{r_{i}-r_{f}}(\tau)][(i\overrightarrow{\nabla}_{r_{f}} +\frac{e\overrightarrow{B}}{2c}\times(\overrightarrow{r_{f}} - \overrightarrow{r_{i}}))^ {\beta} (\delta(-\tau)\delta(r_{f}-r_{i}))] \nonumber\\
 \Pi_ {\alpha,\beta} ^{E,e,4}(r_{i},r_{f};\tau) &=& \frac{ei}{4m^2}[(i\overrightarrow{\nabla}_{r_{i}} +\frac{e\overrightarrow{B}}{2c}\times(\overrightarrow{r_{f}} - \overrightarrow{r_{i}}))^ {\alpha} \widetilde{G}_{r_{f}-r_{i}}(-\tau)][(i\overrightarrow{\nabla}_{r_{f}} -\frac{e\overrightarrow{B}}{2c}\times(\overrightarrow{r_{f}} - \overrightarrow{r_{i}}))^ {\beta}(\frac{\partial \widetilde{G}_{r_{i}-r_{f}}(\tau)}{\partial \tau})] \nonumber\\
 &+& \frac{ei}{4m^2}[(i\overrightarrow{\nabla}_{r_{i}} +\frac{e\overrightarrow{B}}{2c}\times(\overrightarrow{r_{f}} - \overrightarrow{r_{i}}))^ {\alpha} \widetilde{G}_{r_{f}-r_{i}}(-\tau)][(i\overrightarrow{\nabla}_{r_{f}} -\frac{e\overrightarrow{B}}{2c}\times(\overrightarrow{r_{f}} - \overrightarrow{r_{i}}))^ {\beta}(\delta(\tau)\delta(r_{i}-r_{f}))] \nonumber\
\end{eqnarray}
\end{widetext}

\noindent In momentum space, with the definition $\overrightarrow{K}=\overrightarrow{p}+\overrightarrow{q}$, we get

\begin{widetext}
\begin{eqnarray}
 \Pi_ {\alpha,\beta} ^{E,e}(\overrightarrow{q};i\Omega) &=& \Pi_ {\alpha,\beta} ^{E,e,1} (\overrightarrow{q};i\Omega) + \Pi_ {\alpha,\beta}^{E,e,2} (\overrightarrow{q};i\Omega)  + \Pi_ {\alpha,\beta} ^{E,e,3} (\overrightarrow{q};i\Omega)  + \Pi_ {\alpha,\beta} ^{E,e,4} (\overrightarrow{q};i\Omega) \\
 \Pi_ {\alpha,\beta} ^{E,e,1} (\overrightarrow{q};i\Omega) &=&\frac{-ei} {4m^2 \beta}\sum_{\overrightarrow{p},iw_{n}}[(-\overrightarrow{K} -i\frac{e\overrightarrow{B}}{2c}\times \overrightarrow{\nabla}_{\overrightarrow{K}} )^ {\alpha} (\overrightarrow{K} -i\frac{e\overrightarrow{B}}{2c}\times\overrightarrow{\nabla}_{\overrightarrow{K}})^{\beta}\widetilde{G}_{\overrightarrow{K}}(iw_{n} + i\Omega)](-iw_{n} \widetilde{G}_{\overrightarrow{p}}(iw_{n})) \nonumber\\
&+&\frac{ei} {4m^2 \beta}\sum_{\overrightarrow{p},iw_{n}} [(-\overrightarrow{K} -i\frac{e\overrightarrow{B}}{2c}\times \overrightarrow{\nabla}_{\overrightarrow{K}} )^ {\alpha} (\overrightarrow{K} -i\frac{e\overrightarrow{B}}{2c}\times\overrightarrow{\nabla}_{\overrightarrow{K}})^{\beta}\widetilde{G}_{\overrightarrow{K}}(iw_{n} + i\Omega)] \nonumber\\
\Pi_ {\alpha,\beta} ^{E,e,2} (\overrightarrow{q};i\Omega) &=& \frac {ei} {4m^2 \beta}\sum_{\overrightarrow{p},iw_{n}} (iw_{n}+i\Omega) \widetilde{G}_{\overrightarrow{K}}(iw_{n}+i \Omega)[(-\overrightarrow{p} +i\frac{e\overrightarrow{B}}{2c}\times \overrightarrow{\nabla}_{\overrightarrow {p}} ) ^{\alpha} (\overrightarrow{p} +i\frac{e\overrightarrow{B}}{2c}\times\overrightarrow{\nabla}_{\overrightarrow {p}}) ^ {\beta}\widetilde{G}_{\overrightarrow{p}}(iw_{n})] \nonumber\\
&+& \frac {ei} {4m^2 \beta}\sum_{\overrightarrow{p},iw_{n}} [(-\overrightarrow{p} +i\frac{e\overrightarrow{B}}{2c}\times \overrightarrow{\nabla}_{\overrightarrow {p}} ) ^{\alpha} (\overrightarrow{p} +i\frac{e\overrightarrow{B}}{2c}\times\overrightarrow{\nabla}_{\overrightarrow {p}}) ^ {\beta}\widetilde{G}_{\overrightarrow{p}}(iw_{n})] \nonumber\\
 \Pi_ {\alpha,\beta} ^{E,e,3} (\overrightarrow{q};i\Omega) &=& \frac{ei} {4m^2 \beta}\sum_{\overrightarrow{p},iw_{n}}[(-\overrightarrow{K} -i\frac{e\overrightarrow{B}}{2c}\times \overrightarrow{\nabla}_{\overrightarrow{K}} ) ^ {\alpha} \widetilde{G}_{\overrightarrow{K}}(iw_{n}+i \Omega)][(-\overrightarrow{p}-i\frac{e\overrightarrow{B}}{2c}\times\overrightarrow{\nabla}_{\overrightarrow {p}}) ^ {\beta} (-iw_{n} \widetilde{G}_{\overrightarrow{p}}(iw_{n}))] \nonumber\\
 &-& \frac{ei} {4m^2 \beta}\sum_{\overrightarrow{p},iw_{n}} [(-\overrightarrow{K} -i\frac{e\overrightarrow{B}}{2c}\times \overrightarrow{\nabla}_{\overrightarrow{K}} ) ^ {\alpha} \widetilde{G}_{\overrightarrow{K}}(iw_{n}+i \Omega)] [(-\overrightarrow{p}-i\frac{e\overrightarrow{B}}{2c}\times\overrightarrow{\nabla}_{\overrightarrow {p}}) ^ {\beta} (1)] \nonumber\\
 \Pi_ {\alpha,\beta} ^{E,e,4} (\overrightarrow{q};i\Omega) &=& \frac{-ei} {4m^2 \beta}\sum_{\overrightarrow{p},iw_{n}}[(\overrightarrow{p} -i\frac{e\overrightarrow{B}}{2c}\times \overrightarrow{\nabla}_{\overrightarrow {p}} )^ {\alpha} \widetilde{G}_{\overrightarrow{p}}(iw_{n})][(\overrightarrow{K} -i\frac{e\overrightarrow{B}}{2c}\times\overrightarrow{\nabla}_{\overrightarrow{K}}) ^ {\beta}((iw_{n}+i\Omega) \widetilde{G}_{\overrightarrow{K}}(iw_{n}+i \Omega))] \nonumber\\
 &+& \frac{-ei} {4m^2 \beta}\sum_{\overrightarrow{p},iw_{n}} [(\overrightarrow{p} -i\frac{e\overrightarrow{B}}{2c}\times \overrightarrow{\nabla}_{\overrightarrow {p}} )^ {\alpha} \widetilde{G}_{\overrightarrow{p}}(iw_{n})][(\overrightarrow{K} -i\frac{e\overrightarrow{B}}{2c}\times \overrightarrow{\nabla}_{\overrightarrow{K}} ) ^ {\beta}(1)] \nonumber\
\end{eqnarray}
\end{widetext}

\noindent To first order in magnetic field,

\begin{widetext}
\begin{eqnarray}
 \Pi_ {x,x} ^{E,e} (\overrightarrow{q}\to0;i\Omega) &=&\frac{-ei} {\beta}\sum_{\overrightarrow{p},iw_{n}} \{((iw_{n}+\frac{i\Omega}{2}) \overrightarrow{v}_{\overrightarrow{p},x}^{2}) \widetilde{G}_ {\overrightarrow{p}}(iw_{n} + i\Omega)] \widetilde{G}_ {\overrightarrow{p}} (iw_{n}) +\overrightarrow{v}_{\overrightarrow{p},x}^{2} \frac{\widetilde{G}_ {\overrightarrow{p}}(iw_{n} + i\Omega) + \widetilde{G}_ {\overrightarrow{p}}(iw_{n})}{2} \} \nonumber\\
\Pi_ {x,y} ^{E,e} (\overrightarrow{q}\to0;i\Omega) &=&\frac{-e^{2}| \overrightarrow {B}|}{4c\beta m}\sum_{\overrightarrow{p},iw_{n}} (iw_{n}+\frac{i\Omega}{2})(\overrightarrow{v}_{\overrightarrow{p},x})\{\frac{\partial \widetilde{G}_ {\overrightarrow{p}}(iw_{n} + i\Omega)}{\partial p_{x}} \widetilde{G}_ {\overrightarrow{p}}(iw_{n}) - \widetilde{G}_ {\overrightarrow{p}}(iw_{n} + i\Omega) \frac{\partial \widetilde{G}_ {\overrightarrow{p}}(iw_{n})}{\partial p_{x}}\} \nonumber\\
 &+&\frac {-e^{2}| \overrightarrow {B}|} {4c\beta m}\sum_{\overrightarrow{p},iw_{n}} (iw_{n}+\frac{i\Omega}{2})(\overrightarrow{v}_{\overrightarrow{p},y})\{\frac{\partial \widetilde{G}_ {\overrightarrow{p}}(iw_{n} + i\Omega)}{\partial p_{y}} \widetilde{G}_ {\overrightarrow{p}}(iw_{n}) -\widetilde{G}_ {\overrightarrow{p}}(iw_{n} + i\Omega) \frac{\partial \widetilde{G}_ {\overrightarrow{p}}(iw_{n})}{\partial p_{y}}\} \nonumber\\
 &+&\frac {-e^{2}| \overrightarrow {B}|} {4c\beta m}\sum_{\overrightarrow{p},iw_{n}} \{ \frac{i\Omega}{2}\widetilde{G}_ {\overrightarrow{p}}(iw_{n} + i\Omega)\widetilde{G}_ {\overrightarrow{p}}(iw_{n}) -\widetilde{G}_ {\overrightarrow{p}}(iw_{n} + i\Omega) + \widetilde{G}_ {\overrightarrow{p}}(iw_{n})\}\
\end{eqnarray}
\end{widetext}

\noindent In the thermoelectric correlation function, only the terms of the form $(iw_{n}+\frac{i\Omega}{2}) $ contribute to the real part of the thermoelectric conductivity. Performing the sum over frequency and taking the dc limit the thermoelectric conductivity tensor is equal to the electrical conductivity kernel multiplied by energy divided by the electron charge. Thus, the thermoelectric conductivity in the dc limit is

\begin{widetext}
\begin{eqnarray}
\beta_{xx}^{DC} &=& \frac{-N_{s}N_{v}e}{4\pi h} \int d\varepsilon \frac{\partial n_{F}}{\partial \mu}(\varepsilon - \mu)(1+\frac{A^2+B^2}{AB} \arctan{\frac {A}{B}})\\
\beta_{xy}^{DC} &=& \frac{-N_{s}N_{v}e^{2}| \overrightarrow {B}|v_{F}^{2}}{2c\pi}\int \frac{d\varepsilon}{\pi} \frac{\partial n_{F}(\varepsilon)}{\partial \mu} (\varepsilon - \mu)\{ \frac{1}{8AB}(\frac{B^{2}-A^{2}}{B^{2}+A^{2}} - \frac{B^{2}+A^{2}}{2AB} \arctan{\frac{2AB}{B^{2}-A^{2}}}) - \frac{AB}{3(A^{2}+B^{2})^{2}}\}\nonumber\\
&-& \frac{N_{s}N_{v}e^{2}| \overrightarrow {B}|v_{F}^{2}}{2c\pi}\int \frac{d\varepsilon}{\pi} n_{F}(\varepsilon)\{\frac{AB}{3(A^{2}+B^{2})^{2}}\}\
\end{eqnarray}
\end{widetext}

 \section{\label{sec:app-heat}  Heat conductivity tensors }

The energy-energy current correlation function is

\begin{widetext}
\begin{eqnarray}
 \Pi_ {\alpha,\beta} ^{E,E}(r_{i},r_{f};\tau_{i},\tau_{f}) &=& <T_{\tau}(\lim_ { \begin{subarray}{1} r_{i}' \to r_{i} \\ \tau_{i}' \to \tau_{i} \end{subarray}} \frac{i}{2m}[(-i\overrightarrow{\nabla}_{r_{i}'}^{\alpha} -e\overrightarrow{A}_ {r_{i}'}^{\alpha})\frac{\partial}{\partial \tau_{i}} + (-i\overrightarrow{\nabla}_{r_{i}}^{\alpha} +e\overrightarrow{A}_ {r_{i}}^{\alpha})\frac{\partial}{\partial \tau_{i}'}  ]\Psi^{\dagger}(r_{i},\tau_{i})\Psi(r_{i}',\tau_{i}'))\nonumber\\
 & & (\lim_ { \begin{subarray}{1} r_{f}' \to r_{f} \\ \tau_{f}' \to \tau_{f} \end{subarray}} \frac{i}{2m}[(i\overrightarrow{\nabla}_{r_{f}'}^{\beta} -e\overrightarrow{A}_ {r_{f}'}^{\beta})\frac{\partial}{\partial \tau_{f}} + (i\overrightarrow{\nabla}_{r_{f}}^{\beta} +e\overrightarrow{A}_ {r_{f}}^{\beta})\frac{\partial}{\partial \tau_{f}'}  ]\Psi^{\dagger}(r_{f}',\tau_{f}')\Psi(r_{f},\tau_{f})) >
\end{eqnarray}
 \end{widetext}
 
\noindent Using $\frac{\partial}{\partial \tau}$ $G(r_{1},r_{2},\pm\tau)$  $ = \pm\delta(\tau)\delta (r_{1}-r_{2}) +$  $ < T_{\tau} \frac{\partial}{\partial \tau}\Psi(r_{1},\tau) \Psi^{\dagger}(r_{2},0) > $ for first order derivatives in $\tau$ and $\frac{\partial^2}{\partial \tau^2}G(r_{1},r_{2},\tau)$ $ + \delta(r_{1}-r_{2}) \frac{\partial}{\partial \tau} \delta(\tau)$ $ = <T_{\tau} \frac{\partial \Psi(r_{1},\tau_{1})}{\partial \tau_{1}} \frac{\partial \Psi^{\dagger}(r_{2},\tau_{2})}{\partial \tau_{2}}>$ for second order derivatives in $\tau$, 
 
\begin{widetext}
\begin{eqnarray}
 \Pi_ {\alpha,\beta} ^{E,E}(r_{i},r_{f};\tau) &=& \Pi_ {\alpha,\beta} ^{E,E,1}(r_{i},r_{f};\tau) + \Pi_ {\alpha,\beta}^{E,E,2} (r_{i},r_{f};\tau)  + \Pi_ {\alpha,\beta} ^{E,E,3}(r_{i},r_{f};\tau)  + \Pi_ {\alpha,\beta} ^{E,E,4}(r_{i},r_{f};\tau) \\
 \Pi_ {\alpha,\beta} ^{E,E,1}(r_{i},r_{f};\tau) &=&\frac{-1} {4m^2}[(-i\overrightarrow{\nabla}_{r_{i}} +\frac{e\overrightarrow{B}}{2c}\times(\overrightarrow{r_{f}} - \overrightarrow{r_{i}}) )^ {\alpha} (i\overrightarrow{\nabla}_{r_{f}} -\frac{e\overrightarrow{B}}{2c}\times(\overrightarrow{r_{f}} - \overrightarrow{r_{i}}))^{\beta}\widetilde{G}_{r_{i}-r_{f}}(\tau)](\frac{\partial^{2} \widetilde{G}_{r_{f}-r_{i}}(-\tau)}{\partial \tau^{2}}) \nonumber\\
&+&\frac{-1} {4m^2}[(-i\overrightarrow{\nabla}_{r_{i}} +\frac{e\overrightarrow{B}}{2c}\times(\overrightarrow{r_{f}} - \overrightarrow{r_{i}}) )^ {\alpha} (i\overrightarrow{\nabla}_{r_{f}} -\frac{e\overrightarrow{B}}{2c}\times(\overrightarrow{r_{f}} - \overrightarrow{r_{i}}))^{\beta}\widetilde{G}_{r_{i}-r_{f}}(\tau)] (\delta(r_{i}-r_{f}) \frac{\partial \delta(-\tau)}{\partial \tau} ) \nonumber\\
\Pi_ {\alpha,\beta} ^{E,E,2}(r_{i},r_{f};\tau) &=& \frac {-1} {4m^2}(\frac{\partial^{2} \widetilde{G}_{r_{i}-r_{f}}(\tau)}{\partial \tau^{2}})[(-i\overrightarrow{\nabla}_{r_{i}} -\frac{e\overrightarrow{B}}{2c}\times(\overrightarrow{r_{f}} - \overrightarrow{r_{i}}))^{\alpha} (i\overrightarrow{\nabla}_{r_{f}} + \frac{e\overrightarrow{B}}{2c}\times(\overrightarrow{r_{f}} - \overrightarrow{r_{i}}))^ {\beta}\widetilde{G}_{r_{f}-r_{i}}(-\tau) \nonumber\\
&+& \frac {1} {4m^2} (\delta(r_{i}-r_{f}) \frac{\partial \delta(\tau)}{\partial \tau} )[(-i\overrightarrow{\nabla}_{r_{i}} -\frac{e\overrightarrow{B}}{2c}\times(\overrightarrow{r_{f}} - \overrightarrow{r_{i}}))^{\alpha} (i\overrightarrow{\nabla}_{r_{f}} + \frac{e\overrightarrow{B}}{2c}\times(\overrightarrow{r_{f}} - \overrightarrow{r_{i}}))^ {\beta}\widetilde{G}_{r_{f}-r_{i}}(-\tau) \nonumber\\
 \Pi_ {\alpha,\beta} ^{E,E,3}(r_{i},r_{f};\tau) &=& \frac{-1} {4m^2}[(-i\overrightarrow{\nabla}_{r_{i}} +\frac{e\overrightarrow{B}}{2c}\times(\overrightarrow{r_{f}} - \overrightarrow{r_{i}}))^ {\alpha} (\frac{\partial \widetilde{G}_{r_{i}-r_{f}}(\tau)}{\partial \tau})][(i\overrightarrow{\nabla}_{r_{f}} +\frac{e\overrightarrow{B}}{2c}\times(\overrightarrow{r_{f}} - \overrightarrow{r_{i}}))^ {\beta} (\frac{\partial \widetilde{G}_{r_{f}-r_{i}}(-\tau)}{\partial \tau})] \nonumber\\
&+& \frac{1} {4m^2}[(-i\overrightarrow{\nabla}_{r_{i}} +\frac{e\overrightarrow{B}}{2c}\times(\overrightarrow{r_{f}} - \overrightarrow{r_{i}}))^ {\alpha} (\frac{\partial \widetilde{G}_{r_{i}-r_{f}}(\tau)}{\partial \tau})][(i\overrightarrow{\nabla}_{r_{f}} +\frac{e\overrightarrow{B}}{2c}\times(\overrightarrow{r_{f}} - \overrightarrow{r_{i}}))^ {\beta} (\delta(\tau)\delta(r_{i}-r_{f}))] \nonumber\\
&+& \frac{-1} {4m^2}[(-i\overrightarrow{\nabla}_{r_{i}} +\frac{e\overrightarrow{B}}{2c}\times(\overrightarrow{r_{f}} - \overrightarrow{r_{i}}))^ {\alpha} (\delta(\tau)\delta(r_{i}-r_{f}))][(i\overrightarrow{\nabla}_{r_{f}} +\frac{e\overrightarrow{B}}{2c}\times(\overrightarrow{r_{f}} - \overrightarrow{r_{i}}))^ {\beta} (\frac{\partial \widetilde{G}_{r_{f}-r_{i}}(-\tau)}{\partial \tau})] \nonumber\\
&+& \frac{1} {4m^2}[(-i\overrightarrow{\nabla}_{r_{i}} +\frac{e\overrightarrow{B}}{2c}\times(\overrightarrow{r_{f}} - \overrightarrow{r_{i}}))^ {\alpha} (\delta(\tau)\delta(r_{i}-r_{f}))][(i\overrightarrow{\nabla}_{r_{f}} +\frac{e\overrightarrow{B}}{2c}\times(\overrightarrow{r_{f}} - \overrightarrow{r_{i}}))^ {\beta} (\delta(\tau)\delta(r_{i}-r_{f}))] \nonumber\\
 \Pi_ {\alpha,\beta} ^{E,E,4}(r_{i},r_{f};\tau) &=& \frac{-1}{4m^2}[(-i\overrightarrow{\nabla}_{r_{i}} -\frac{e\overrightarrow{B}}{2c}\times(\overrightarrow{r_{f}} - \overrightarrow{r_{i}}))^ {\alpha} (\frac{\partial \widetilde{G}_{r_{f}-r_{i}}(-\tau)}{\partial \tau}) ][(i\overrightarrow{\nabla}_{r_{f}} -\frac{e\overrightarrow{B}}{2c}\times(\overrightarrow{r_{f}} - \overrightarrow{r_{i}}))^ {\beta} (\frac{\partial \widetilde{G}_{r_{i}-r_{f}}(\tau)}{\partial \tau})] \nonumber\\
  &+& \frac{-1}{4m^2}[(-i\overrightarrow{\nabla}_{r_{i}} -\frac{e\overrightarrow{B}}{2c}\times(\overrightarrow{r_{f}} - \overrightarrow{r_{i}}))^ {\alpha} (\frac{\partial \widetilde{G}_{r_{f}-r_{i}}(-\tau)}{\partial \tau}) ][(i\overrightarrow{\nabla}_{r_{f}} -\frac{e\overrightarrow{B}}{2c}\times(\overrightarrow{r_{f}} - \overrightarrow{r_{i}}))^ {\beta} (\delta(\tau)\delta(r_{i}-r_{f}))] \nonumber\\
&+& \frac{1}{4m^2}[(-i\overrightarrow{\nabla}_{r_{i}} -\frac{e\overrightarrow{B}}{2c}\times(\overrightarrow{r_{f}} - \overrightarrow{r_{i}}))^ {\alpha} (\delta(-\tau)\delta(r_{f}-r_{i})) ][(i\overrightarrow{\nabla}_{r_{f}} -\frac{e\overrightarrow{B}}{2c}\times(\overrightarrow{r_{f}} - \overrightarrow{r_{i}}))^ {\beta} (\frac{\partial \widetilde{G}_{r_{i}-r_{f}}(\tau)}{\partial \tau})] \nonumber\\
&+& \frac{1}{4m^2}[(-i\overrightarrow{\nabla}_{r_{i}} -\frac{e\overrightarrow{B}}{2c}\times(\overrightarrow{r_{f}} - \overrightarrow{r_{i}}))^ {\alpha} (\delta(-\tau)\delta(r_{f}-r_{i})) ][(i\overrightarrow{\nabla}_{r_{f}} -\frac{e\overrightarrow{B}}{2c}\times(\overrightarrow{r_{f}} - \overrightarrow{r_{i}}))^ {\beta} (\delta(\tau)\delta(r_{i}-r_{f}))] \nonumber\
\end{eqnarray}
\end{widetext}

\noindent In momentum space, with the definition $\overrightarrow{K}=\overrightarrow{p}+\overrightarrow{q}$, we get

\begin{widetext}
\begin{eqnarray}
 \Pi_ {\alpha,\beta} ^{E,E} (\overrightarrow{q};i\Omega) &=& \Pi_ {\alpha,\beta} ^{E,E,1} (\overrightarrow{q};i\Omega) + \Pi_ {\alpha,\beta}^{E,E,2} (\overrightarrow{q};i\Omega)  + \Pi_ {\alpha,\beta} ^{E,E,3} (\overrightarrow{q};i\Omega)  + \Pi_ {\alpha,\beta} ^{E,E,4} (\overrightarrow{q};i\Omega) \\
 \Pi_ {\alpha,\beta} ^{E,E,1} (\overrightarrow{q};i\Omega) &=&\frac{1}{4m^2\beta}\sum_{\overrightarrow{p},iw_{n}}[(\overrightarrow{K} +\frac{ie\overrightarrow{B}}{2c}\times \overrightarrow{\nabla}_{\overrightarrow{K}} )^ {\alpha} (\overrightarrow{K} -\frac{ie\overrightarrow{B}}{2c}\times \overrightarrow{\nabla}_{\overrightarrow{K}} ) ^{\beta}(\widetilde{G}_{\overrightarrow{K}}(iw_{n}+i\Omega))]((iw_{n})^{2} \widetilde{G}_{\overrightarrow{p}}(iw_{n})) \nonumber\\
&+&\frac{1}{4m^2\beta}\sum_{\overrightarrow{p},iw_{n}}[(\overrightarrow{K} +\frac{ie\overrightarrow{B}}{2c}\times \overrightarrow{\nabla}_{\overrightarrow{K}} )^ {\alpha} (\overrightarrow{K} -\frac{ie\overrightarrow{B}}{2c}\times \overrightarrow{\nabla}_{\overrightarrow{K}} ) ^{\beta}(\widetilde{G}_{\overrightarrow{K}}(iw_{n}+i\Omega))] (-iw_{n}) \nonumber\\
\Pi_ {\alpha,\beta} ^{E,E,2} (\overrightarrow{q};i\Omega) &=&\frac{1}{4m^2\beta}\sum_{\overrightarrow{p},iw_{n}} ((iw_{n}+i\Omega) ^{2} \widetilde{G}_{\overrightarrow{K}}(iw_{n}+i\Omega))[(-\overrightarrow{p} +\frac{ie\overrightarrow{B}}{2c}\times\overrightarrow{\nabla}_{\overrightarrow{p}})^{\alpha} (-\overrightarrow{p} -\frac{ie\overrightarrow{B}}{2c}\times\overrightarrow{\nabla}_{\overrightarrow{p}}) ^ {\beta}(\widetilde{G}_{\overrightarrow{p}}(iw_{n}))]\nonumber\\
&-&\frac{1}{4m^2\beta}\sum_{\overrightarrow{p},iw_{n}} (iw_{n}+i\Omega)[(-\overrightarrow{p} +\frac{ie\overrightarrow{B}}{2c}\times\overrightarrow{\nabla}_{\overrightarrow{p}})^{\alpha} (-\overrightarrow{p} -\frac{ie\overrightarrow{B}}{2c}\times\overrightarrow{\nabla}{\overrightarrow{p}}) ^ {\beta}(\widetilde{G}_{\overrightarrow{p}}(iw_{n}))]\nonumber\\
\Pi_ {\alpha,\beta} ^{E,E,3} (\overrightarrow{q};i\Omega) &=&\frac{1}{4m^2\beta}\sum_{\overrightarrow{p},iw_{n}} [(\overrightarrow{K} +\frac{ie\overrightarrow{B}}{2c}\times \overrightarrow{\nabla}_{\overrightarrow{K}})^ {\alpha} ((iw_{n}+i\Omega)\widetilde{G}_{\overrightarrow{K}}(iw_{n}+i\Omega))][(-\overrightarrow{p} -\frac{ie\overrightarrow{B}}{2c}\times \overrightarrow{\nabla}_{\overrightarrow{p}}) ^ {\beta} ((-iw_{n})\widetilde{G}_{\overrightarrow{p}}(iw_{n}))] \nonumber\\
&-&\frac{1}{4m^2\beta}\sum_{\overrightarrow{p},iw_{n}} [(\overrightarrow{K} +\frac{ie\overrightarrow{B}}{2c}\times \overrightarrow{\nabla}_{\overrightarrow{K}})^ {\alpha} ((iw_{n}+i\Omega)\widetilde{G}_{\overrightarrow{K}}(iw_{n}+i\Omega))][(-\overrightarrow{p} -\frac{ie\overrightarrow{B}}{2c}\times \overrightarrow{\nabla}_{\overrightarrow{p}}) ^ {\beta} (1)] \nonumber\\
&+&\frac{1}{4m^2\beta}\sum_{\overrightarrow{p},iw_{n}} [(\overrightarrow{K} +\frac{ie\overrightarrow{B}}{2c}\times \overrightarrow{\nabla}_{\overrightarrow{K}})^ {\alpha} (1)][(-\overrightarrow{p} -\frac{ie\overrightarrow{B}}{2c}\times \overrightarrow{\nabla}_{\overrightarrow{p}}) ^ {\beta} ((-iw_{n})\widetilde{G}_{\overrightarrow{p}}(iw_{n}))] \nonumber\\
&-&\frac{1}{4m^2\beta}\sum_{\overrightarrow{p},iw_{n}} [(\overrightarrow{K} +\frac{ie\overrightarrow{B}}{2c}\times \overrightarrow{\nabla}_{\overrightarrow{K}})^ {\alpha} (1)][(-\overrightarrow{p} -\frac{ie\overrightarrow{B}}{2c}\times \overrightarrow{\nabla}_{\overrightarrow{p}}) ^ {\beta} (1)] \nonumber\\
\Pi_ {\alpha,\beta} ^{E,E,4} (\overrightarrow{q};i\Omega) &=&\frac{1}{4m^2\beta}\sum_{\overrightarrow{p},iw_{n}} [(-\overrightarrow{p} + \frac{ie\overrightarrow{B}}{2c}\times \overrightarrow{\nabla}_{\overrightarrow{p}})^ {\alpha} ((-iw_{n})\widetilde{G}_{\overrightarrow{p}}(iw_{n}))][(\overrightarrow{K} -\frac{ie\overrightarrow{B}}{2c}\times \overrightarrow{\nabla}_{\overrightarrow{K}}) ^ {\beta} ((iw_{n} + i\Omega)\widetilde{G}_{\overrightarrow{p}}(iw_{n}+i \Omega))] \nonumber\\
&+&\frac{1}{4m^2\beta}\sum_{\overrightarrow{p},iw_{n}} [(-\overrightarrow{p} + \frac{ie\overrightarrow{B}}{2c}\times \overrightarrow{\nabla}_{\overrightarrow{p}})^ {\alpha} ((-iw_{n})\widetilde{G}_{\overrightarrow{p}}(iw_{n}))][(\overrightarrow{K} -\frac{ie\overrightarrow{B}}{2c}\times \overrightarrow{\nabla}_{\overrightarrow{K}}) ^ {\beta} (1)] \nonumber\\
&-&\frac{1}{4m^2\beta}\sum_{\overrightarrow{p},iw_{n}} [(-\overrightarrow{p} + \frac{ie\overrightarrow{B}}{2c}\times \overrightarrow{\nabla}_{\overrightarrow{p}})^ {\alpha} (1)][(\overrightarrow{K} -\frac{ie\overrightarrow{B}}{2c}\times \overrightarrow{\nabla}_{\overrightarrow{K}}) ^ {\beta} ((iw_{n} + i\Omega)\widetilde{G}_{\overrightarrow{p}}(iw_{n}+i \Omega))] \nonumber\\
&-&\frac{1}{4m^2\beta}\sum_{\overrightarrow{p},iw_{n}} [(-\overrightarrow{p} + \frac{ie\overrightarrow{B}}{2c}\times \overrightarrow{\nabla}_{\overrightarrow{p}})^ {\alpha} (1)][(\overrightarrow{K} -\frac{ie\overrightarrow{B}}{2c}\times \overrightarrow{\nabla}_{\overrightarrow{K}}) ^ {\beta} (1)] \nonumber\
\end{eqnarray}
\end{widetext}

\noindent To first order in magnetic field

\begin{widetext}
\begin{eqnarray}
 \Pi_ {x,x} ^{E,E}(\overrightarrow{q}\to0;i\Omega) &=&\frac{1} {\beta}\sum_{\overrightarrow{p},iw_{n}} \{(iw_{n}+\frac{i\Omega}{2})^{2} \overrightarrow{v}_{\overrightarrow{p},x}^{2} \widetilde{G}(\overrightarrow{p}, iw_{n} + i\Omega) \widetilde{G} (\overrightarrow{p}, iw_{n}) + \frac{\overrightarrow{v}_{\overrightarrow{p},x}^{2}}{2}\} \nonumber\\
 &+& \frac{1} {\beta}\sum_{\overrightarrow{p},iw_{n}} \{\frac{\overrightarrow{v}_{\overrightarrow{p},x}^{2}}{2}\frac{(iw_{n}+i\Omega)\widetilde{G}(\overrightarrow{p}, iw_{n} + i\Omega) + iw_{n}\widetilde{G} (\overrightarrow{p}, iw_{n})}{2} + \frac{i\Omega \overrightarrow{v}_{\overrightarrow{p},x}^{2}}{2}\frac{\widetilde{G}(\overrightarrow{p}, iw_{n} + i\Omega) -\widetilde{G} (\overrightarrow{p}, iw_{n})}{2} \} \nonumber\\
\Pi_ {x,y} ^{E,E} (\overrightarrow{q}\to0;i\Omega) &=&\frac{-ie| \overrightarrow {B}|}{4c\beta m}\sum_{\overrightarrow{p},iw_{n}} (iw_{n}+\frac{i\Omega}{2})^{2}(\overrightarrow{v}_{\overrightarrow{p},x})\{\frac{\partial \widetilde{G}_ {\overrightarrow{p}}(iw_{n} + i\Omega)}{\partial p_{x}} \widetilde{G}_ {\overrightarrow{p}}(iw_{n}) - \widetilde{G}_ {\overrightarrow{p}}(iw_{n} + i\Omega) \frac{\partial \widetilde{G}_ {\overrightarrow{p}}(iw_{n})}{\partial p_{x}}\} \nonumber\\
 &+&\frac {-ie| \overrightarrow {B}|} {4c\beta m}\sum_{\overrightarrow{p},iw_{n}} (iw_{n}+\frac{i\Omega}{2})^{2}(\overrightarrow{v}_{\overrightarrow{p},y})\{\frac{\partial \widetilde{G}_ {\overrightarrow{p}}(iw_{n} + i\Omega)}{\partial p_{y}} \widetilde{G}_ {\overrightarrow{p}}(iw_{n}) -\widetilde{G}_ {\overrightarrow{p}}(iw_{n} + i\Omega) \frac{\partial \widetilde{G}_ {\overrightarrow{p}}(iw_{n})}{\partial p_{y}}\} \nonumber\\
 &+&\frac {-ie| \overrightarrow {B}|} {4c\beta m}\sum_{\overrightarrow{p},iw_{n}} (iw_{n}+\frac{i\Omega}{2})\{ \frac{i\Omega}{2}\widetilde{G}_ {\overrightarrow{p}}(iw_{n} + i\Omega)\widetilde{G}_ {\overrightarrow{p}}(iw_{n}) - \widetilde{G}_ {\overrightarrow{p}}(iw_{n} + i\Omega) + \widetilde{G}_ {\overrightarrow{p}}(iw_{n})\}\
\end{eqnarray}
\end{widetext}

\noindent By performing the sum over frequency and taking the dc limit of the energy conductivity tensor, the result is just the electrical conductivity kernel multiplied by energy divided by the electron charge squared. Energy conductivity in the dc limit is

\begin{widetext}
\begin{eqnarray}
\kappa_{xx}^{DC} &=& \frac{N_{s}N_{v}}{4\pi h} \int d\varepsilon \frac{\partial n_{F}}{\partial \mu}(\varepsilon - \mu)^{2}(1+\frac{A^2+B^2}{AB} \arctan{\frac {A}{B}})\\
\kappa_{xy}^{DC} &=& \frac{-N_{s}N_{v}e| \overrightarrow {B}|v_{F}^{2}}{2c\pi}\int \frac{d\varepsilon}{\pi} \frac{\partial n_{F}(\varepsilon)}{\partial \mu} (\varepsilon - \mu)^{2}\{ \frac{1}{8AB}(\frac{B^{2}-A^{2}}{B^{2}+A^{2}} - \frac{B^{2}+A^{2}}{2AB} \arctan{\frac{2AB}{B^{2}-A^{2}}}) - \frac{AB}{3(A^{2}+B^{2})^{2}}\}\nonumber\\
&-& \frac{N_{s}N_{v}e| \overrightarrow {B}|v_{F}^{2}}{c\pi}\int \frac{d\varepsilon}{\pi} n_{F}(\varepsilon) (\varepsilon - \mu)\{\frac{AB}{3(A^{2}+B^{2})^{2}}\}\
\end{eqnarray}
\end{widetext}
\bibliographystyle {unsrt}
\bibliography{references}

\begin{thebibliography}{10}

\bibitem{Novoselov11102005}
K.~S. Novoselov.
\newblock {\em Nature}, 438(10):197--200, Nov 2005.

\bibitem{PhysRevLett.99.246803}
Y.-W. Tan, Y.~Zhang, K.~Bolotin, Y.~Zhao, S.~Adam, E.~H. Hwang, S.~Das~Sarma,
  H.~L. Stormer, and P.~Kim.
\newblock {\em Phys. Rev. Lett.}, 99(24):246803, Dec 2007.

\bibitem{NatMater362007}
A.~K. Geim and K.~S. Novoselov.
\newblock {\em Nat Mater}, 6(2):183--191, Mar 2007.

\bibitem{Katsnelson200720}
M.I. Katsnelson.
\newblock {\em Materials Today}, 10(1-2):20 -- 27, 2007.

\bibitem{NatPhys822006}
G.~Katsnelson, Novoselov K.S., and A.~Geim.
\newblock {\em Nat Phys}, 2(20):620--625, Aug 2006.

\bibitem{NatPhys212009}
A.~F. Young and P.~Kim.
\newblock {\em Nat Phys}, 5(1):222--226, Feb 2009.

\bibitem{PhysRevLett.102.026807}
N.~Stander, B.~Huard, and D.~Goldhaber-Gordon.
\newblock {\em Phys. Rev. Lett.}, 102(2):026807, Jan 2009.

\bibitem{Novoselov02152007}
K.~S. Novoselov.
\newblock {\em Science}, page 1137201, 2007.

\bibitem{Nature872009}
X.~Du, I.~Skachko, F.~Duerr, A.~Luican, and E.~Y. Andrei.
\newblock {\em Nature}, 462(7):192--195, Aug 2009.

\bibitem{PhysRevLett.102.166808}
P.~Wei, W.~Bao, Y.~Pu, C.N. Lau, and J~Shi.
\newblock {\em Phys. Rev. Lett.}, 102(16):166808, Apr 2009.

\bibitem{PhysRevLett.102.096807}
Y.~M. Zuev, W.~Chang, and P.~Kim.
\newblock {\em Phys. Rev. Lett.}, 102(9):096807, Mar 2009.

\bibitem{PhysRevB.80.081413}
J.~G. Checkelsky and N.~P. Ong.
\newblock {\em Phys. Rev. B}, 80(8):081413, Aug 2009.

\bibitem{Alexander-A.-Balandin:2008hc}
A.A. Balandin, S.~Ghosh, W.~Bao, I.~Calizo, D.~Teweldebrhan, F.~Miao, and C.N.
  Lau.
\newblock {\em Nano Lett.}, 8(3):902--907, Feb 2008.

\bibitem{Adam20091072}
S.~Adam, E.~Hwang, E.H.and~Rossi, and S.~Das~Sarma.
\newblock {\em Solid State Communications}, 149(27-28):1072 -- 1079, 2009.
\newblock Recent Progress in Graphene Studies.

\bibitem{2010arXiv1003.4731D}
S.~{Das Sarma}, S.~{Adam}, E.~H. {Hwang}, and E.~{Rossi}.
\newblock {\em ArXiv e-prints}, mar 2010.

\bibitem{Wallace:1947lr}
P.~R. Wallace.
\newblock {\em Phys. Rev.}, 71(9):622--634, May 1947.

\bibitem{PhysRevB.79.125427}
Cr. Bena.
\newblock {\em Phys. Rev. B}, 79(12):125427, Mar 2009.

\bibitem{PhysRevB.78.085416}
L.~Fritz, J.~Schmalian, M.~Muller, and S.~Sachdev.
\newblock {\em Phys. Rev. B}, 78(8):085416, Aug 2008.

\bibitem{PhysRevLett.98.186806}
E.~H. Hwang, S.~Adam, and S.~Das~Sarma.
\newblock {\em Phys. Rev. Lett.}, 98(18):186806, May 2007.

\bibitem{JPSJ.67.2421}
N-H Shon and T.~Ando.
\newblock {\em Journal of the Physical Society of Japan}, 67(7):2421--2429,
  1998.

\bibitem{JPSJ.71.1318}
T.~Ando, Y.~Zheng, and H.~Suzuura.
\newblock {\em Journal of the Physical Society of Japan}, 71(5):1318--1324,
  2002.

\bibitem{PhysRevB.73.245403}
M.~Koshino and T.~Ando.
\newblock {\em Phys. Rev. B}, 73(24):245403, Jun 2006.

\bibitem{PhysRevB.73.125411}
N.~M.~R. Peres, F.~Guinea, and A.~H. Castro~Neto.
\newblock {\em Phys. Rev. B}, 73(12):125411, Mar 2006.

\bibitem{PhysRevB.77.125409}
X-Z Yan, Y.~Romiah, and C.~S. Ting.
\newblock {\em Phys. Rev. B}, 77(12):125409, Mar 2008.

\bibitem{Katsnelson:2006fr}
M.I. Katsnelson.
\newblock {\em Euro. Phys. Jour. B}, 51(157):157--160, May 2006.

\bibitem{PhysRevLett.104.076804}
L.~Zhu, R.~Ma, L.~Sheng, M.~Liu, and D-N Sheng.
\newblock {\em Phys. Rev. Lett.}, 104(7):076804, Feb 2010.

\bibitem{NaturePhys542008}
J.H. Chen, C.~Jang, S.~Adam, M.S. Fuhrer, E.D. Williams, and M.~Ihigami.
\newblock {\em Nat Phys}, 4(5):377--381, May 2008.

\bibitem{NatPhys02042008}
J.~Martin, N.~Akerman, G.~Ulbricht, T.~Lohmann, J.~H. Smet, K.~von Klitzing,
  and A.~Yacoby.
\newblock {\em Nat Phys}, 4(2):144--148, Feb 2008.

\bibitem{wang:043121}
Y.~Y. Wang, Z.~H. Ni, Z.~X. Shen, H.~M. Wang, and Y.~H. Wu.
\newblock {\em Applied Physics Letters}, 92(4):043121, 2008.

\bibitem{ACS030332009}
Z.~H. Ni, T.~Yu, Z.~Q. Luo, Ying~Y. Wang, L.~Liu, C.~P. Wong, J.~Miao,
  W.~Huang, and Z.~X. Shen.
\newblock {\em ACS Nano}, 3(3):569--574, 03 2009.

\bibitem{schmittrink}
S.~Schmitt-Rink, K.~Miyake, and C.M. Varma.
\newblock {\em Phys. Rev. Lett}, 57:2575, 1986.

\bibitem{aleiner}
I.~L. Aleiner and K.~B. Efetov.
\newblock {\em Phys. Rev. Lett.}, 97(23):236801, Dec 2006.

\bibitem{khvesch}
Khveshchenko.
\newblock {\em Euro. Phys. Lett.}, 827.

\bibitem{peresrmp}
N.~M.~R. Peres.
\newblock {\em Rev. Mod. Phys.}, 82(3):2673--2700, Sep 2010.

\bibitem{PhysRevB.76.193401}
T.~Lofwander and M.~Fogelstrom.
\newblock {\em Phys. Rev. B}, 76(19):193401, Nov 2007.

\bibitem{ashcroft}
Neil~W. Ashcroft and David~N. Mermin.
\newblock Thomson Learning, Toronto, 1 edition, January 1976.

\bibitem{2007PNAS..10418392A}
S.~{Adam}, E.~H. {Hwang}, V.~M. {Galitski}, and S.~{Das Sarma}.
\newblock {\em Proceedings of the National Academy of Science},
  104:18392--18397, Nov 2007.

\bibitem{polini}
Marco Polini, Andrea Tomadin, Reza Asgari, and A.~H. MacDonald.
\newblock {\em Phys. Rev. B}, 78(11):115426, Sep 2008.

\bibitem{nla.cat-vn1437043}
G.~D. Mahan.
\newblock Plenum Press, New York :, 1981.

\bibitem{PhysRevB.68.155114}
M.~Khodas and A.~M. Finkel'stein.
\newblock {\em Phys. Rev. B}, 68(15):155114, Oct 2003.

\bibitem{PhysRevB.67.115131}
I.~Paul and G.~Kotliar.
\newblock {\em Phys. Rev. B}, 67(11):115131, Mar 2003.

\bibitem{JPSJ.76.043711}
H.~Fukuyama.
\newblock {\em Journal of the Physical Society of Japan}, 76(4):043711, 2007.

\bibitem{PhysRevB.73.245411}
V.~P. Gusynin and S.~G. Sharapov.
\newblock {\em Phys. Rev. B}, 73(24):245411, Jun 2006.

\bibitem{PhysRevB.31.7291}
H.~Oji and P.~Streda.
\newblock {\em Phys. Rev. B}, 31(11):7291--7295, Jun 1985.

\bibitem{PhysRevB.55.2344}
N.~R. Cooper, B.~I. Halperin, and I.~M. Ruzin.
\newblock {\em Phys. Rev. B}, 55(4):2344--2359, Jan 1997.

\bibitem{Smrcka-Streda:1977SS}
L.~Smrcka and P.~Streda.
\newblock {\em J. Phys. C: Solid State Phys.}, 10(2153):2153--2161, Jun 1977.

\bibitem{PhysRevB.80.235415}
E.~H. Hwang, E.~Rossi, and S.~Das~Sarma.
\newblock {\em Phys. Rev. B}, 80(23):235415, Dec 2009.

\bibitem{PhysRevB.80.155423}
X-Z. Yan and C.~S. Ting.
\newblock {\em Phys. Rev. B}, 80(15):155423, Oct 2009.

\bibitem{PhysRevB.80.165423}
X-Z. Yan, Y.~Romiah, and C.~S. Ting.
\newblock {\em Phys. Rev. B}, 80(16):165423, Oct 2009.

\bibitem{PhysRevB.81.155457}
X-Z Yan and C.~S. Ting.
\newblock {\em Phys. Rev. B}, 81(15):155457, Apr 2010.

\end{thebibliography}

\end{document}